\documentclass[journal]{IEEEtran}

\usepackage{rotating}
\usepackage{subfigure}
\usepackage{graphicx}

\usepackage{booktabs}		
\usepackage{longtable}
\usepackage{array}				
 		
\usepackage{mathtools}
\usepackage{amsmath}
\usepackage{amssymb}

\usepackage{url}

\usepackage{lscape}

\newcommand{\bitem}[1]{\item{\textbf{#1}}:}
\newcommand{\tablespace}{\noalign{\smallskip}}
\newtheorem{definition}{Definition}
\newcommand{\botrule}{\bottomrule}
\newcommand{\colrule}{\midrule}

\begin{document}

\title{On Representing Resilience Requirements of Microservice Architecture Systems}

\author{Kanglin Yin, Qingfeng Du}

\maketitle

\begin{abstract}
Together with the spread of DevOps practices and container technologies, Microserivce Architecture has become a mainstream architecture style in recent years.
Resilience is a key characteristic in Microservice Architecture Systems(MSA Systems), and it shows the ability to cope with various kinds of system disturbances which cause degradations of services.
However, due to lack of consensus definition of resilience in the software field, although a lot of work has been done on resilience for MSA Systems, 
developers still don't have a clear idea on how resilient an MSA System should be, 
and what resilience mechanisms are needed.  

In this paper, 
by referring to existing systematic studies on resilience in other scientific areas, 
the definition of microservice resilience is provided and a Microservice Resilience Measurement Model is proposed to measure service resilience.
And a requirement model to represent resilience requirements of MSA Systems is given.
The requirement model uses elements in KAOS to represent notions in the measurement model, and decompose service resilience goals into system behaviors that can be executed by system components.
As a proof of concept, a case study is conducted on an MSA System to illustrate how the proposed models are applied.
\end{abstract}


\footnotetext[1]{This manuscript is draft only, not intended for publication}

\section{Introduction}
\label{sec:introduction}


Microservice Architecture (aka Microservices)\cite{fowler2014microservices} is a new architectural style which modularizes software components as services, that are called microservices,
and these services are fine-grained and isolated so that they are able to change independently of each other.
In recent years, Microservice Architecture(MSA) has become a mainstream architecture style adopted by many leading internet companies\cite{hoff2017lessons}\cite{mauro2016adopting}.
Shifting to Microservice Architecture enables development with fast iterations\cite{newman2015building},
therefore developers can adopt modern development processes like Continuous Delivery\cite{humble2010continuous}, DevOps\cite{bass2015devops} to ensure time-to-market services \cite{balalaie2016microservices}.


Microservices are usually deployed in a sophisticated environment using virtual infrastructure like virtual machines and containers, and there are lots of components for service decoupling and management (e.g. API Gateway, Message Queue, Service Registry, etc.). 
Thus threats come from anywhere in an MSA System: 
small-density components with faults \cite{hatton1997reexamining}, 
unstable message passing among microservices\cite{montesi2016circuit}, 
the underlying cloud environment with unreliable containers, virtual machines,  or bare-metal servers, etc.\cite{esposito2016challenges}.
Even normal actions taken in cloud environments like software/hardware upgrades and dynamic changes in configuration files may lead to severe service outages, 
which are lessons learned from hundreds of service outage reports of cloud companies in the literature\cite{Gunawi2016Why}.

Reliability, availability, fault tolerance etc. are traditional system properties to evaluate a software system's ability to cope with failures \cite{organizacion2011iso}. 
The metrics of these properties(like MTBF) usually assume that a software system has two states such as "Available/Unavailable" or "Reliable/Unreliable", 
and calculate the probability of these two states.
However, recent studies on cloud system failures \cite{Gunawi2016Why}\cite{Gunawi2014What}\cite{gunawi2018fail} found that cloud systems are more likely to be in a "limped" state rather than be totally unavailable.
The "limped" state\cite{gunawi2018fail} means that although a cloud system can provide services with normal functionalities, the services work in performance under users' satisfaction, which is known as service degradation.
In such situation, metrics of reliability or availability can't evaluate the software system so well.
For example, 
suppose two cases occurred in an MSA system:
in case A, the average response time of an MSA System's service is degraded from 3 seconds to 5 seconds due to a failure; while in case B, the average response time of the service is degraded from 3 seconds to 12 seconds due to the same failure.
The failure recovery time of these two cases are the same.
It is obvious that the service in case A performs better than the service in case B when failure happens.
But if we take "response time higher than 3 seconds" as an unreliable state, the metrics of service reliability in these two cases are the same.


As a result, many MSA practitioners \cite{newman2015building}\cite{wolff2016microservices}\cite{nadareishvili2016microservice} 
proposed \textbf{Resilience} as a characteristic describing how an MSA System copes with the occurrence of failures and recovers the degraded service back to its normal performance.
Although engineering resilience in MSA Systems has gained popularity among designers and engineers,
until now there is no common definition for microservice resilience.
Several works have been done for software resilience benchmarking \cite{bondavalli2009research}\cite{almeida2012changeloads}, but available engineering quantification metrics still exhibit very little standardization.
Due to no standard definitions and quantification metrics for microservice resilience, it is hard to make definite resilience requirements for MSA Systems.
As a result, microservice developers seldom have a clear idea of the following questions about resilience:
\begin{enumerate}
\item What is microservice resilience?
\item How to evaluate microservice resilience?
\item How resilient an MSA System should be?
\end{enumerate}

Furthermore, how to represent resilience requirements for MSA Systems is an another problem to face even if standard definitions and quantification metrics of microservice resilience are given.
Although Domain Driven Design (DDD) \cite{evans2004domain} is the suggested way to build MSA System requirements \cite{newman2015building},
integrating resilience into DDD seems to be difficult.
DDD focuses on how to decompose a system to microservices by business context boundaries,
so it doesn't comprise service deployment components like containers, or notions like service degradations and failures\cite{DBLP:journals/software/RademacherSS18}.
Another type of requirement model is needed for microservice resilience requirements.

In order to solve the problems above, this paper works on the representation and elicitation of MSA System resilience requirements, contributions of this paper are:
\begin{itemize}
\item 
We provide the definition of microservice resilience. 
And a Microservice Resilience Measurement Model (MRMM) is proposed to measure service resilience of MSA Systems.
\item Based on MRMM, a requirement model is designed to represent resilience requirements of MSA Systems. 
The requirement model contains a Resilience Goal Decomposition View refining service resilience goals to system behaviors with a customized goal model,
and a Resilience Mechanism Implementation View to show how resilience mechanisms work in MSA Systems.
\end{itemize}

The remain of this paper organizes as follows:
Section \ref{sec:related} summarises some related works of this paper;
Section \ref{sec:resilience} provides the definition of microservice resilience and proposed Microservice Resilience Measurement Model;
Section \ref{sec:representation} proposes the service resilience requirement model based on the definition and measurement model in Section \ref{sec:resilience};
In Section \ref{sec:case} a case study is conducted using an MSA System to illustrate the propose resilience requirement model,
and Section \ref{sec:conclusion} makes conclusion of this paper and outlines some future works.

\section{Related Works}
\label{sec:related}

This section discusses existing related studies in these three research areas: Resilience in other scientific areas,
Resilience in microservices,
and Goal-Oriented Requirement Engineering.

\subsection{Resilience in Other Scientific Areas}
\label{sec:related-res}
The word "resilience"
means an object's ability to bonus back to its normal size after being pulled or pushed.
Holling \cite{holling1973resilience} firstly used this word in the field of ecology, to represent an ecosystem's ability to absorb disturbances.
In recent decades, the notion of resilience has been used in many scientific areas, like psychology, city planning, management science, etc.
As is described in Righi's review of resilience engineering\cite{righi2015systematic}, the ability of anticipating/being aware of hazards, the capacity of adapting to variability, and the ability of responding/restoring are major concerns of resilience.

In Hosseini's literature review on system resilience\cite{hosseini2016review},
resilience assessment is classified into two categories: qualitative approaches and quantitative approaches.
Qualitative approaches gives concept framework of practices to archieve resilience\cite{ungar2003qualitative}, which is usually used in Society-Ecology, Organization Management, and Healthcare.
Quantitative approaches use \textbf{resilience curves} to illustrate the resilient behavior of an engineered system undergoing a \textbf{disruption}.
Many researchers used the properties of the resilience curve to measure the resilience of the system.
In this paper quantitative approaches are used to measure the resilience of MSA systems. 

\begin{figure}
\centering
\includegraphics[width = 0.7\linewidth]{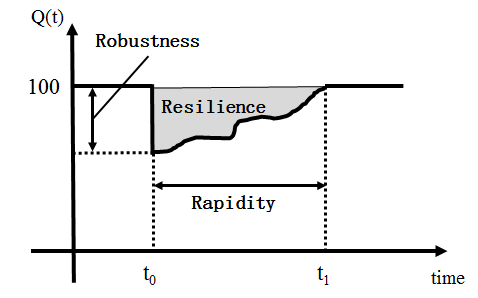}
\caption{The Bruneau's Reslience Triangle Model}
\label{fig:resilience-tri}
\end{figure}

Bruneau's resilience triangle model\cite{bruneau2003framework} is the most used resilience model in quantitative assessment.
Based on the resilience curve illustrating a system's quality variation under a disruption event, the Bruneau's model proposed three dimensions of Resilience: \textit{Robustness}, \textit{Rapidity} and \textit{Resilience}, as is shown in Figure \ref{fig:resilience-tri}. 
In Figure \ref{fig:resilience-tri}, the x-axis stands for the time and the y-axis stands for the quality of a system.
The cumulated performance loss of a system is represented as the shaded part in Figure \ref{fig:resilience-tri}.
The \textit{Resilience} dimension measures the area of this shaded part, 
and the other two dimensions measure the shaded part's projection length on the x and y axis.

A great number of researches on resilience quantification has been done based on the resilience curve.
Based on Bruneau's model, these works simplified calculation methods, customized existing models, or proposed derived metrics from three dimensions of resilience according to the need of different research areas, as is concluded by reviews of resilience quantification\cite{henry2012generic}\cite{yodo2016engineering}\cite{hosseini2016review}.
The resilience model proposed in this paper will also refer to metrics in Bruneau's model to set resilience goal of services.

\subsection{Resilience in Microservices}
\label{sec:related-ms}
The importance of resilience has been pointed out in many practitioner books of microservices \cite{newman2015building}\cite{wolff2016microservices}\cite{nadareishvili2016microservice}, and studies discussing key features of MSA Systems \cite{dragoni2017microservices}\cite{dragoni2017microserviceshow}. 
In these books and studies, 
some typical system mechanisms consisting of operations that react to service degradations are mentioned, 
such as Load Balancing, Circuit Breakers, API Gateway, Bulkheads, etc.
These mechanisms are called resilience mechanisms.
Some researchers improved these resilience mechanisms for MSA Systems in recent years.\cite{Brilhante:2017:AQB:3126858.3126873}\cite{7965189}\cite{8486300}\cite{8530769}\cite{Montesi:2018:DPC:3167132.3167427}

Besides resilience mechanisms, there are several other resilience related works on MSA Systems.
Richter et al. showed that the Microservice Architecture itself can have positive impacts on dependability and fault-tolerance\cite{richter2017highly}.
Nane analyzed the impact of containers to microservice performance \cite{kratzke2017microservices}.
Giovanni, et al, proposed a self-managing mechanism for MSA System\cite{toffetti2015architecture}, where auto-scaling and health management of services are implemented by monitoring. 
Soenen et al. designed a scalable mechanism for microservice-based NFV System for fault recovery and high availability \cite{soenen2017optimising}.
Zwietasch used Time-Series Prediction method to predict failures in MSA Systems\cite{zwietasch2017online}.
Stefan et al. designed a decision guidance model for service discovery and fault of microservices, where some faults in an MSA System are related to certain designs in the system \cite{haselbock2017decision}.
Heorhiadi et al, designed a resilience testing framework for the Microservice Architecture \cite{heorhiadi2016gremlin}, how to inject faults into microservices was not discussed in detail.
Brogi proposed a reference dataset generation framework for microservices which includes failure data\cite{brogi2017towards}.
Thomas and Andre, built a meta-model for MSA Systems, which is used for performance and resilience benchmarking\cite{D2017Model}, but how this model is used for benchmarking and how resilience is evaluated was not discussed.

Above works also pointed out that resilience is important for MSA System, and find facts or give solutions on resilience design and resilience testing.
However, none of these works discussed about resilience requirements of MSA Systems.
Neither the definition nor the measurement of microservice resilience are given.
And none of research questions in Section \ref{sec:introduction} can be answered through these works.

\subsection{Goal-Oriented Requirement Engineering}
\label{sec:related-gore}

Goal-oriented Requirement Engineering (GORE) is a branch of requirement engineering.
GORE is used in earlier stages of requirements analysis to elicit the requirements from high-level system goals\cite{van2001goal}, while object-oriented analysis like UML\cite{rubin1992object} fits well to the later stages.
\textit{KAOS}\cite{van2004object}\cite{dardenne1993goal}, \textit{i*}\cite{yu1997towards}, \textit{Tropos}\cite{DBLP:journals/aamas/BrescianiPGGM04}, \textit{GBRAM}\cite{anton1996goal} and the \textit{NFR Framework}\cite{mylopoulos1992representing} are main goal modeling frameworks used in GORE\cite{lapouchnian2005goal}.
In this paper, a requirement model based on KAOS is proposed. It consists of basic elements in KAOS, including:
\begin{itemize}
\item \textbf{goal}: objective the system should meet;
\item \textbf{refinement}: relationship linking a goal to other goals that are called its \textbf{subgoals}. Each subgoal contributes to the satisfaction of the goal it refines;
\item \textbf{obstacle}: condition whose satisfaction may prevent goal(s) from being achieved;
\item \textbf{agent}: active object that acts as processor for some operations;
\item \textbf{terminal goal} (or called \textbf{requirement}): goal assignable to individual agent, it's where goal refinement ends up;
\item \textbf{domain property}: assertion about objects in the environment of the system.
\end{itemize}

On GORE with failures or performance,
Rashid et al. extended the goal-oriented model to aspect-oriented model, which separates the functional and non-functional properties of a requirement \cite{rashid2002early}.
Van and Letier integrated the notion of \textbf{obstacle}, which obstructs the goal satisfication, into their KAOS model in \cite{van1998integrating}.
And then Axel presented formal techniques for reasoning about obstacles to satisfaction of goals in his paper \cite{Van2000Handling}, which mainly focuses on finding and resolving obstacles in functional goals that are too ideal to be realized by software.
Van Lamsweerde explicitly modeled the goals of an attacker, an agent that creates security risks for the system goals in KAOS\cite{van2004elaborating}.
John, Lawrence and Brian proposed a goal-oriented framework for Non-Functional Requirements \cite{mylopoulos1992representing}, a series of refinement methods were designed for accuracy and performance requirements.

Few works were done on GORE for microservice-related systems.
Wang et al. introduced service discovery in GORE, business goals in KAOS was decomposed into subgoals of services in SOA system \cite{wang2015discovering}.
Zardari and Bahsoon used GORE to decompose cloud migration goals into services that cloud platforms provide when assessing whether cloud platforms is suited for the target system\cite{Zardari2011Cloud}.
Chung\cite{DBLP:journals/jss/ChungHLSDS13}, Gonçalves\cite{DBLP:conf/sac/JuniorRSM15} and Patil\cite{patil2015cloud} set SLA(Service Level Agreement) thresholds for non-functional requirements of cloud services.
Deprez et al.\cite{deprez2012integrating} built mapping from quantitative, non-functional requirements in an SOA system to detailed UML design.
Duboc et al.\cite{DBLP:journals/tse/DubocLR13} used the notion of goal and obstacle to analyze scalability problems in cloud environment.
Zardari\cite{DBLP:conf/sac/ZardariBE14}\cite{zardari2013using} also used obstacle to represent risks in cloud environments, and used goals to represent solution to these risks. 
The resilience requirement model proposed in this paper will follow the ideas in these studies. 

\section{Microservice Resilience}
\label{sec:resilience}

The definitions of microservice resilience and Microservice Reslience Measurement Model are proposed in this section,
as the basis of resilience requirement representation for MSA Systems.  

\subsection{Definition of Microservice Resilience}
\label{sec:resilience-def}
According to definitions of resilience in different research areas, and what is required to be achieved in MSA Systems, we have found that some resilience-related facts in MSA Systems are consistent with some consensuses on resilience.
Finally we have come to the following conclusions on MSA System resilience:

\begin{itemize}

\item In running environments of MSA Systems, there a lot of unpredictable events that make services of MSA Systems perform not as good as expected.
These events are termed as "disruptions" in the field of resilience\cite{hosseini2016review} \cite{righi2015systematic}.

\item 
It is hard to get the probabilities of disruptions because the architecture and deployment environment of MSA Systems always change with the quick iteration of DevOps.
In other research areas, it is usually assumed that disruptions are inevitable and always happen \cite{hosseini2016review} \cite{yodo2016engineering}. 
Considering high fault density in MSA Systems\cite{dragoni2017microservices},
it can be assumed that disruptions are deterministic events in MSA Systems.

\item 
Service performance is the main concern of MSA System Resilience.
Disruptions in MSA Systems cause losses of service performance, which are called service degradations.
The curve representing how a service's performance varies from time under a service degradation is a typical "resilience curve" in researches of resilience\cite{yodo2016engineering}.

\item 
Resilient MSA Systems should keep performance degraded services from a too low level which is unacceptable by users,
and make degraded services' performance back to normal as fast as possible.

\end{itemize}

Different scientific areas have different emphasis on resilience\cite{hosseini2016review}.
In order to clarify what system characteristics that microservice resilience concerns about,
the definition of resilience for MSA Systems is provided based on the conclusions above:   

\textit{
\textbf{Resilience} of a Microservice Architecture System is the ability to maintain the performance of services at an acceptable level and recover the service back to normal, when a disruption causes the service degradation.
}

\subsection{Microservice Resilinece Measurement Model}
\label{sec:resilience-mrmm}

A variety of resilience metrics on have been proposed in existing works on resilience quantification. These metrics belong to different research areas and have different quantification targets.
In order to quantify resilience of MSA Systems,
we proposed Microservice Resilience Measurement Model (MRMM).
Compared with general resilience models, a concept model is added in MRMM besides the resilience metrics.
The concept model consists of key concepts in MSA System related with resilience, like service, performance, service degradations, etc.
These concepts are used to clarify what is measured by microservice resilience, and will be represented as elements in the proposed resilience requirement model in Section \ref{sec:representation}.
Similar to existing resilience models on quantifying disruption impacts, 
three metrics are used to quantify service degradations occurred in MSA Systems from different dimensions.

\begin{figure}
\centering
\includegraphics[width = \linewidth]{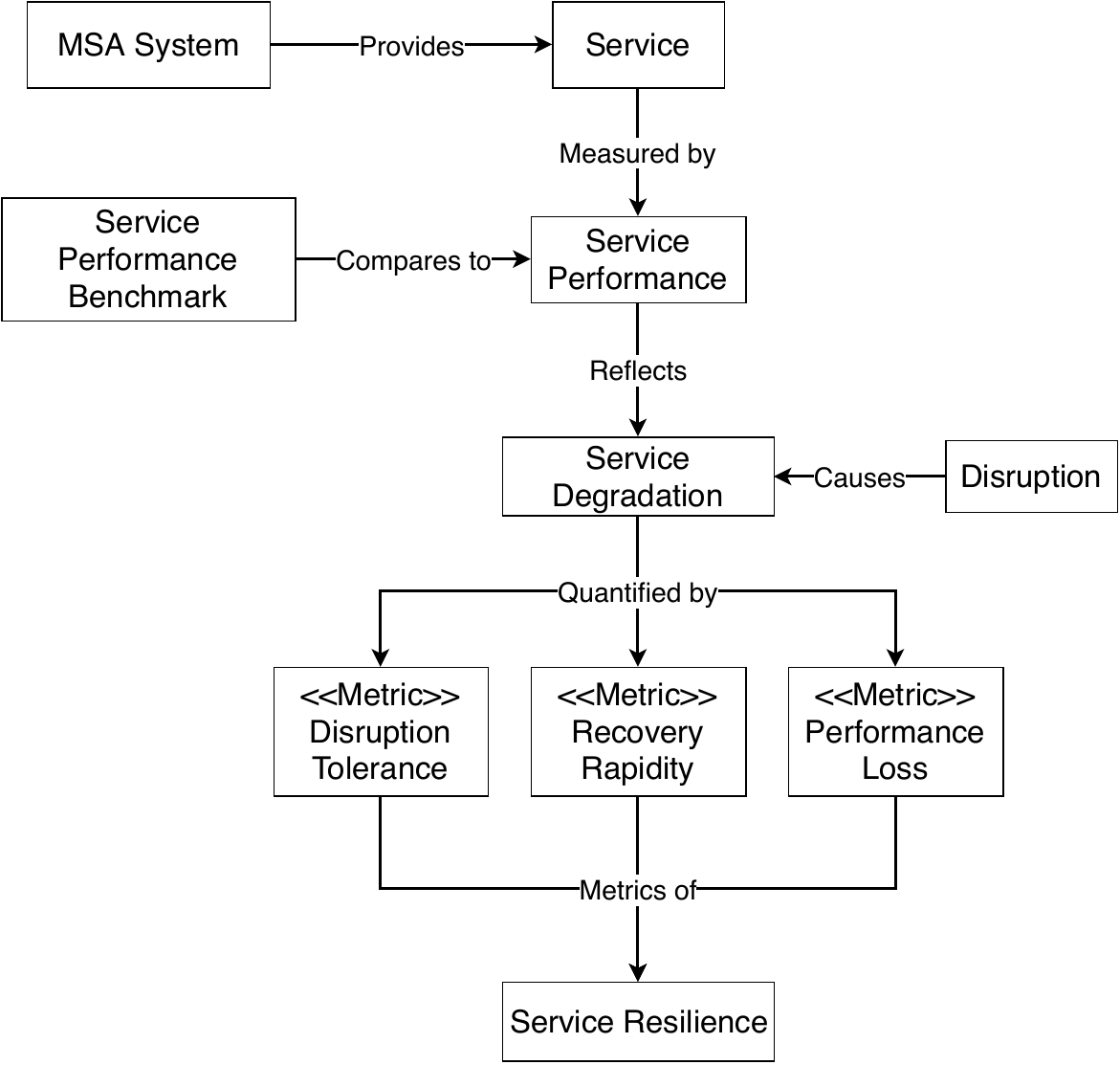}
\caption{Meta-Model of MRMM}
\label{fig:mrmm-meta}
\end{figure}

Figure \ref{fig:mrmm-meta} shows the meta-model of MRMM.
Below we provide the definitions for elements in MRMM.
Mathematical presentations of these elements are also given, in order to encode resilience requirements to formal propositions and verify the satisfiability of resilience requirements in our future work.

\begin{definition}[MSA System and Service]

An \textbf{MSA System} is a software system that provides  \textbf{services}.
Every service in an MSA System is an interface exposed to users or other systems.
Users and other systems can fulfil certain functionalities by accessing services.
Factors like how service are modularized and deployed are not included in our definition for MSA Systems, because they are out of the scope of our definition on microservice resilience.

In mathematics, An MSA System is represented by a set $MS=\{S_1, \dots , S_n\}$ where $S_1, \dots , S_n$ are services provided by the MSA System.
Each service is represented by a tuple $ S = \langle L, \mathcal{A} \rangle $ where:

\begin{itemize}

\item $L$ is the identification label of the service,
which is used for verification in Goal Models\cite{Giorgini2002Reasoning}.

\item  $ \mathcal{A} = \{ A_1, \dots, A_n \}$ is the set of performance attributes of the service. 
Performance attribute is defined in the next paragraph.

\end{itemize}

\end{definition}

\bigskip

\begin{definition}[Service Performance]

Each service in an MSA System has one or several \textbf{performance attributes} (e.g. response time, throughput) to evaluate service performance.
Performance Attributes of a service are decided by type of the service.
For example, availability and success rate are common performance attributes of transactional services, while video stream services are usually benchmarked by throughputs.

\textbf{Service performance} is metric of a service's performance attribute in a certain time period. 
Service performance can be represented by function $P(S,A,t)$, where $S$ is the service, $A$ is the performance attribute, and $t$ is the timestamp.

In well-developed MSA Systems, real-time service performance data can be collected by monitoring tools like cAdvisor, Zabbix.
These data are stored in time-series databases so that the performance value of a service at a timestamp can be queried in the form of $P(S,A,t)$. 

\end{definition}

\bigskip

\begin{definition}[Service Performance Benchmark]

A \textbf{service performance benchmark} is the baseline value of service performance.
Depending on the types of services and performance attributes, a service performance benchmark may be either a constant value, 
or a dynamic value varying from time. 
Service performance benchmarks are used to judge whether services are degraded.
If the service performance value is lower than its service performance benchmark at some time, the corresponding service is regarded as degraded.

Same with the mathematical representation of service performance, service performance benchmark is represented by function $Q_B(S,A,t)$,
meaning the baseline value of service $S$'s performance attribute $a$ at time $t$.
In mathematics, if the predicate $Degrad(S,t)$ means a service $S$ is degraded at time t, $Degrad(S,t)$ can be defined by the following propositional formula.
\begin{equation}
\label{eq:pre-degrad}
Degrad(S,t) \longleftrightarrow \exists I, P(S,I,t) < P_B(S,I,t)
\end{equation}

\end{definition}

\bigskip

\begin{definition}[Disruption]

A \textbf{disruption} in an MSA System is an event that happens to the MSA System which makes a service degraded. A disruption should contain the following information:

\begin{itemize}
\item The related objects when a disruption happens. 
An object may be any abstract or realistic entity that can be identified in an MSA system (e.g. servers, containers, CPUs, processes, network connections, services).
\item The event type of a disruption. 
For an object where a disruption happens, there are several event types. 
For example, a Virtual Machine's disruption event type may include VM halt down, OS upgrade, kernel break, etc.
\end{itemize}

A disruption is represented by the tuple $D=\langle O, E \rangle$ in mathematics, where $O$,$E$ are the labels of the object and event type.
If the predicate $Occur(D,t)$ means a disruption $D$ occurs at timestamp $t$, the fact that $D$ causes service degradation on service $S$ at timestamp $t'$ can be presented by Eq.(\ref{eq:pre-dis}).

\begin{equation}
\label{eq:pre-dis}
Occur(D, t) \longrightarrow Degrad(S, t')
\end{equation}

\end{definition}

\bigskip

\begin{definition}[Service Degradation and Service Resilience]

\textbf{Service degradation} is a phenomenon happening in MSA Systems that a service is kept degraded because of a disruption.
Degraded state of a service is confirmed by judging whether the service performance is lower than the service performance benchmark.
In mathematics, a service degradation is represented by tuple $SD=\langle S, A, D, t_s, t_e \rangle$, where:
\begin{itemize}
\item $S$ is the degraded service;
\item $A$ is the performance attribute where service performance benchmark is violated;
\item $D$ is the disruption causing the service degradation;
\item $t_s$ and $t_e$ are the start time and the end time of the service degradation.
\end{itemize}

\textbf{Service resilience} is a concept that represents the impact of a service degradation.
Since service performance values under a service degradation is a typical resilience curve, 
three metrics are used to quantify service resilience from different dimensions of the resilience curve:
Disruption Tolerance, Recovery Rapidity and Performance Loss.
Performance Loss refers to Bruneau's resilience triangle model\cite{bruneau2003framework} which is the most used resilience quantification metric. 
Disruption Tolerance and Recovery Rapidity are common metrics derived from a resilience curve in existing resilience quantifications models.
Different from systems in other research areas where these metrics are used, in MSA Systems resilience curves reflect service degradations, which is a unique notion in MSA Systems.
And service degradations are judged by service performance benchmarks which may be dynamic values varying from time. 
While in systems of other research areas, the steady-state performance value is usually constant.
So the definition of these metrics are customized according to definitions of other elements in MRMM.


\begin{itemize}

\bitem{Disruption Tolerance}

\textbf{Disruption Tolerance} measures how much service performance is degraded compared with service performance benchmark.
Disruption Tolerance of a service degradation is the maximum deviation of service performance from service performance benchmark in the period of service degradation.
In mathematics, the Disruption Tolerance $DT(SD)$ of a service degradation $SD$ is represented by Eq.(\ref{eq:dis-tor}).

\begin{multline}
\label{eq:dis-tor}
DT(SD) = \max( P_B(SD.S,SD.A,t) \\
- P(SD.S,SD.A,t) ), t \in [SD.t_s, SD.t_e]
\end{multline}

When a service is suffering degradation, the MSA System should keep the service from severe degradation which is unacceptable to users (For example, the frame rate of a video stream service can be lowered a bit but not too low to make a video look like a slide).

\bigskip

\bitem{Recovery Rapidity}

\textbf{Recovery Rapidity} measures how fast a degraded service can be recovered and reach the service performance benchmark again.
Similar to Mean Time to Repair (MTTR) used in reliability assessment, Recovery Rapidity $RR(SD)$ is measured by calculating the time interval of the service degradation $SD$, as is shown in Eq.(\ref{eq:rec-rapid}).

\begin{equation}
\label{eq:rec-rapid}
RR(SD) = SD.t_e - SD.t_s
\end{equation}

\bigskip

\bitem{Performance Loss}

\textbf{Performance Loss} is a quantification of the magnitude of service degradation in service performance.
Performance Loss $PL(SD)$ of a service degradation $SD$ is mathematically expressed by Eq.(\ref{eq:qual-loss}).

\begin{multline}
\label{eq:qual-loss}
PL(SD) = \int_{SD.t_s}^{SD.t_e} [P_B(SD.S,SD.A,t) \\
- P(SD.S,SD.A,t)] dt
\end{multline}

Performance Loss measures the cumulative degraded performance during the service degradation, which is shown as the shaded area in Figure \ref{fig:resilience-tri}.
Performance Loss can reveal business loss in a service degradation.
For example, if a data transmission service benchmarked by throughput suffers a service degradation, Performance Loss can measure how much data is less transmitted than expected.

In time series databases used for monitoring, there already exist data types used for recording cumulative values of performance (Like the Counter data type in Prometheus).
So it is possible to collect Performance Loss data of service degradations in MSA Systems.

\end{itemize}

\end{definition}

\begin{figure}
\centering
\includegraphics[width = .6\linewidth]{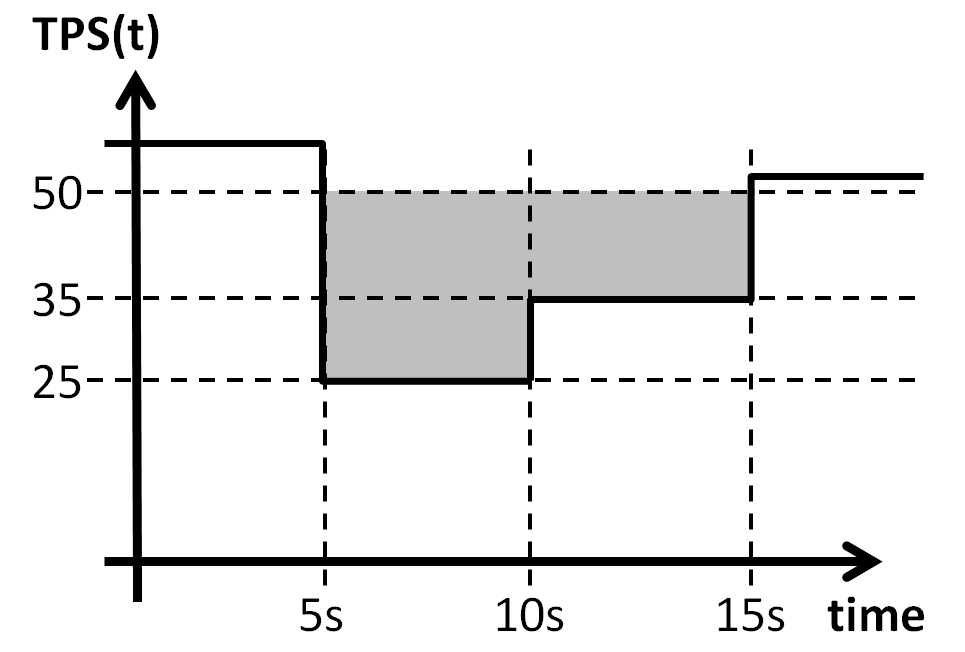}
\caption{Performance on TPS of a Sample Service Degradation}
\label{fig:resilience-sample}
\end{figure}

Service resilience can measure a service degradation with these three metrics.
Mathematically, Service resilience is represented by tuple $SR=\langle DT, RR, PL \rangle$, where $DT$, $RR$  and $PL$ are Disruption Tolerance, Recovery Rapidity, and Performance Loss.
Take a service with performance attribute TPS (Transactions Per Second) as an example. 
The performance benchmark on TPS of this service is 50 requests/second.
When the service suffered from a disruption and was recovered later, and collected TPS value during service degradation is shown in Figure \ref{fig:resilience-sample}.
Disruption Tolerance of the service is $ 50-25 = requests/second$, 
Recovery Rapidity of the service is $15-5=10$ seconds, 
and Performance Loss is $5*(50-35)+5*(50-25) = 200$ requests, which means 200 user requests are less processed than expected due to the disruption.

\section{Resilience Requirement Representation}
\label{sec:representation}

In Section \ref{sec:resilience}, we proposed MRMM to characterize service resilience.
With service resilience metrics, microservice practitioners can set service resilience goals to indicate how resilient an MSA System is supposed to be.
Service resilience goals specify service resilience with thresholds of service resilience metrics in MRMM.
For a service with a service resilience goal, 
any service degradation happened to this service shall happen within the thresholds of the service resilience goal.

When there is a service degradation that violates the service resilience goal, the service degradation should be further analyzed to diagnose the root cause disruption.
Then developers establish corresponding resilience mechanism to mitigate the impact of the identified disruption so that the service resilience goal is satisfied again.
In MSA Systems, a resilience mechanism is a process consisting of system behaviors (like monitoring, failure detection) executed by one or several components to react to disruptions.
And microservice practitioners usually use architecture-level diagrams(like the module view and the component-and-connector view in literature\cite{clements2002documenting}) to show how a resilience mechanism works in MSA Systems\cite{newman2015building}\cite{Hole2016Anti}.

Thus the resilience requirement of an MSA System consists of the following information based on our definition on microservice resilience and MRMM:
\begin{itemize}
\item Service resilience goals;
\item Disruptions that cause service resilience goal violations;
\item Resilience mechanisms established to mitigate impacts of disruptions.
\end{itemize}

We proposed Service Resilience Requirement Model to represent the microservice resilience requirement above.
The proposed requirement model consists of two views: Resilience Goal Decomposition View and Resilience Mechanism Implementation View.
The \textbf{Resilience Goal Decomposition View} is a customized model of a mainstream goal model, KAOS\cite{dardenne1993goal}, which is more expressive than other goal models\cite{nguyen2018multi}, 
and notions in MRMM are integrated into this goal model.
The model uses basic elements in KAOS: \textbf{Goal}, \textbf{Obstacle}, \textbf{Agent} and \textbf{Domain Property}.
Besides these elements, we add an new element: \textbf{Asset},
which represents services and system resources in MSA Systems.
The \textbf{Resilience Mechanism Implementation View} uses microservice practitioners' existing documentation styles for resilience mechanisms (like architectural-level diagrams in \cite{newman2015building}\cite{Hole2016Anti}), 
to show how resilience mechanisms work in MSA Systems in a more expressive way.

\begin{sidewaysfigure*}
\centering
\includegraphics[width = \linewidth]{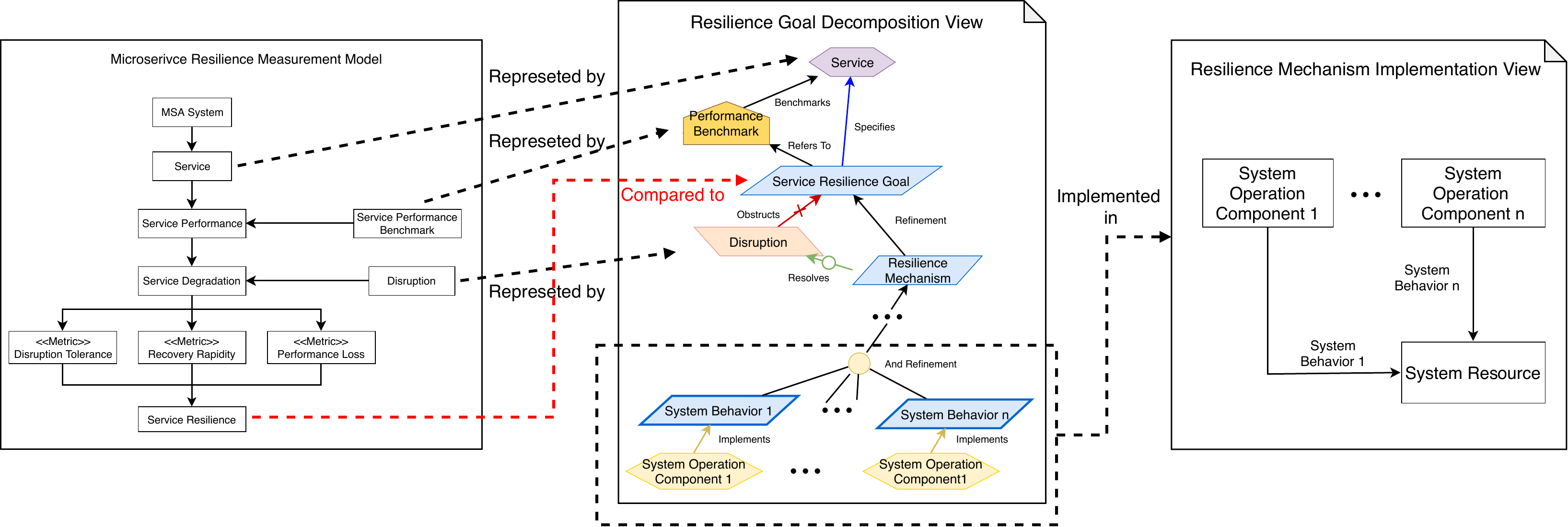}
\caption{Relations between MRMM and Service Resilience Requirement Model}
\label{fig:re-model}
\end{sidewaysfigure*}

In Resilience Goal Decomposition View, service resilience goals are set and decomposed with the methodology of Goal-Oriented Requirement Engineering (GORE) \cite{van2001goal}:
Service resilience goals are decomposed into resource resilience goals(e.g. resilience goals of services' containers or number of available service instances).
Then service degradations that violate service resilience goals  are identified.
Service degradations are further analyzed and root cause disruptions that obstructs resource resilience goals are found out.
In order to resolve disruptions, resilience mechanism are established.
Finally detailed system behaviors are functionally decomposed from resilience mechanisms. These system behaviors will be executed by components of MSA Systems.
The decomposed system behaviors are intertwined with the elements in Resilience Mechanism Implementation View.
Figure \ref{fig:re-model} shows the symbolic representations of the elements in the proposed model, and the model's relation with MRMM.

h


\subsection{Resilience Goal Decomposition View}
In Resilience Goal Decomposition View, service resilience goals are set and decomposed with the methodology of Goal-Oriented Requirement Engineering (GORE) \cite{van2001goal}:
service resilience goals are decomposed into resource resilience goals,
disruptions obstructing service/resource resilience goals are identified,
resilience mechanisms resolving obstacles are established,
and resilience mechanisms are further refined to detailed system behaviors which can be implemented by components of MSA Systems.
The following paragraphs explains in detail how resilience requirements of an MSA Systems are represented in Resilience Goal Decomposition View,
and some mathematical representations are given, for our future work on verifying resilience goal satisfaction.

\

\subsubsection{Service Resilience Goal}

\textbf{Goals} are objectives that the target system should achieve. 
In KAOS, goals cover different types of concerns: from high-level, strategic concerns, to low-level, technical concerns; from functional concerns, to non-functional concerns\cite{van2001goal}.
In diagrams of our proposed goal model, goals are represented by blue parallelograms.

\textbf{Service resilience goals} are final goals to be achieved in an MSA System, which specify how resilient the services in MSA Systems are supposed to be.
A service resilience goal of a service contains thresholds of service resilience metrics in MRMM.
And the performance attribute is also specified in a service resilience goal, 
because service resilience metrics are calculated from service performance variations.

Mathematically, 
if $RG(S,A)$ means the service resilience goal $RG$ of service $S$ on performance attribute $A$, 
and the predicate $satisfy(RG)$ means $RG$ is satisfied,
based on definitions of service degradation and service resilience in MRMM,  
$RG$ satisfies the propositional formula in Eq.(\ref{eq:satis-rg}), where $\mathcal{SD}(S,A)$ means all service degradations of service $S$ on performance attribute $A$.

\begin{multline}
\label{eq:satis-rg}
satisfy(RG(S,A)) \longleftrightarrow 
\forall SD \in \mathcal{SD}(S,A), \\
(RE(SD).DT < RG.DT) \wedge \\
(RE(SD).RR < RG.RR) \wedge \\
(RE(SD).PL < RG.PL)
\end{multline}

\begin{figure}
\centering
\includegraphics[width = .7\linewidth]{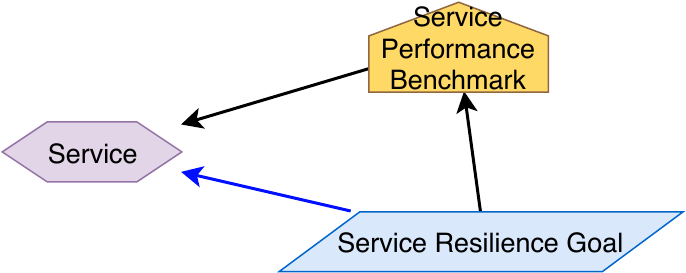}
\caption{The Service Resilience Goal of a Service benchmarked by a Service Performance Attribute}
\label{fig:svc-res-goal}
\end{figure}

A service resilience goal links an asset representing a service with the \textbf{specify} link, and a domain property representing the service performance benchmark with the \textbf{refers to} link, 
to show which service the service resilience goal specifies and which service performance benchmark is used to calculate service resilience when service degradation happens, 
as is shown in Figure \ref{fig:svc-res-goal}.

In KAOS models, textual specification of elements is required\cite{darimont1997grail}.
We use the following information to specify a service resilience goal:
\begin{itemize}
\item \textbf{Goal Name}: The identifier of the service resilience goal;
\item \textbf{Service}: The service which the service resilience goal specifies;
\item \textbf{Performance Attribute}: The performance attribute of the service resilience goal;
\bitem{Service Resilience Thresholds} Thresholds of service resilience metrics.
\end{itemize}

\bigskip

\subsubsection{Resilience Goal Refinement}

\

In KAOS model, A high-level goal can be refined to low-level goals, these low-level goals are called \textbf{sub goals} of the high-level goal.
There are two types of refinements: AND-refinement and OR-refinement.

\textbf{AND-refinement} means a goal can be achieved by satisfying all of its sub goals. Given a goal $G_0$ and a set of $G_0$'s sub goals $\{G_1, \dots, G_n \}$, AND-refinement is textually denoted as $ G_0 \xrightarrow{AND} \{G_1,\dots,G_n\} $.
Mathematically, AND-refinement satisfies the propositional Formula \ref{eq:and-refinement}.

\begin{equation}
\label{eq:and-refinement}
\bigwedge_{i=1}^{n} satisfied(G_i)
\longrightarrow satisfied(G_0)
\end{equation}

\textbf{OR-refinement} means a goal can be achieved when one of its sub goals is satisfied.
Given a goal $G_0$ and a set of $G_0$'s sub goals $\{G_1, \dots, G_n \}$, OR-refinement is textually denoted as $ G_0 \xrightarrow{OR} \{G_1,\dots,G_n\} $.
Mathematically, OR Refinement satisfies the propositional Formula \ref{eq:or-refinement}.

\begin{equation}
\label{eq:or-refinement}
\bigvee_{i=1}^{n} satisfied(G_i)
\longrightarrow satisfied(G_0)
\end{equation}

\begin{figure}[htbp]
  \centering
  \subfigure[And Refinement]{
      \centering
      \includegraphics[width=.7\linewidth]{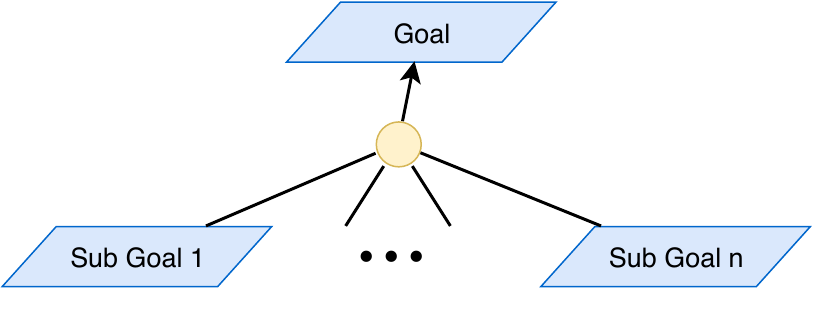}
  }
  \subfigure[OR Refinement]{
      \centering
      \includegraphics[width=.7\linewidth]{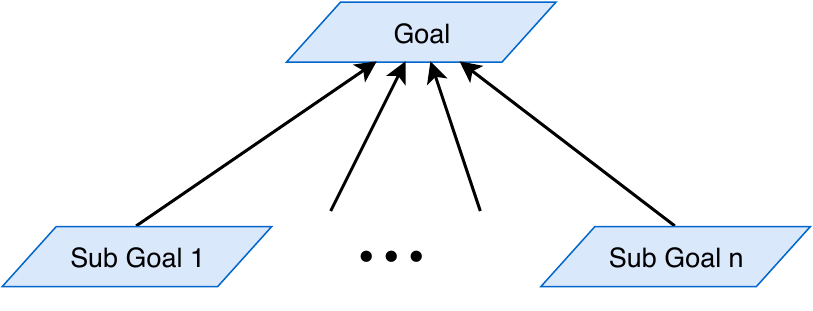}
  }
  \caption{And Refinement and OR Refinement}
  \label{fig:refinement}
\end{figure}

Figure \ref{fig:refinement} shows the difference in representation between AND-refinement and OR-refinement in Resilience Goal Decomposition View.  

In our goal model,
each service resilience goal is refined to resilience goals of service's dependency system resources which ensure the running of the service (such as containers, VMs), if the performance attribute of the service resilience goal is directly influenced by performance attributes of these system resources.
Resource resilience goals can be further refined to resilience goals of resources' dependency system resources.
Such refinement is an AND-refinement because a service resilience goal may be obstructed by just one resource resilience goal violation.

When disruptions that obstruct service/resource resilience goals are found, resilience mechanisms are established to resolve disruptions.
These resilience mechanisms are sub goals of service/resource resilience goals since they promise the satisfaction of service/resource resilience goals.

Resilience mechanisms are functionally decomposed, 
until detailed system behaviors which can be executed by individual components in MSA Systems are figured out.
System behaviors of resilience mechanisms are represented by blue parallelograms with bold borders to show that they are \textbf{terminal goals}.
Each system behavior is linked with an agent representing a system component, to show which component executes the system behavior.

\begin{figure}
\centering
\includegraphics[width = \linewidth]{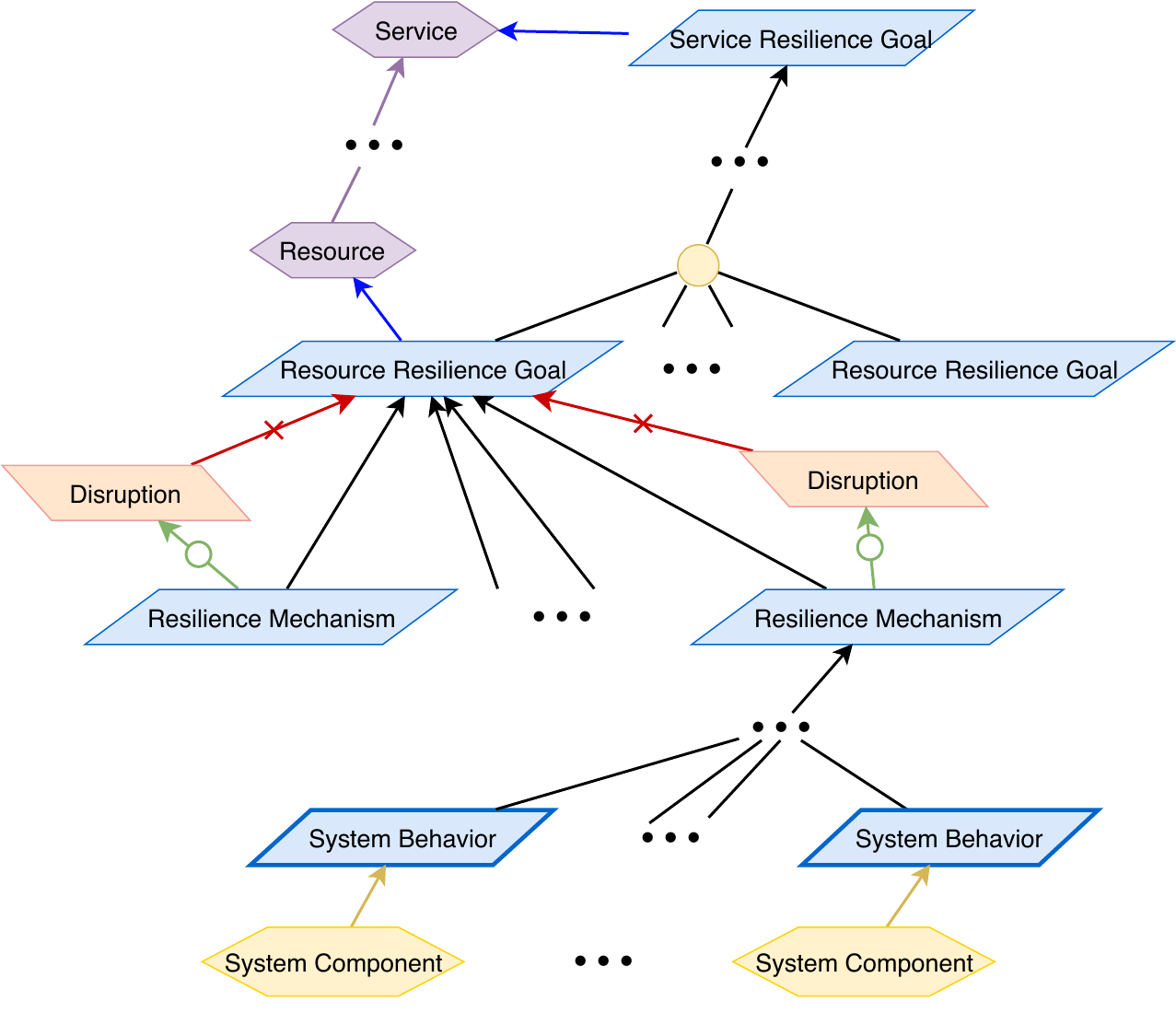}
\caption{Goal Refinement from a Service Resilience Goal to System Behaviors}
\label{fig:total-refinement}
\end{figure}

Figure \ref{fig:total-refinement} shows the whole goal refinement from a service resilience goal to system behaviors.

\bigskip

\subsubsection{Resilience Obstacle}
\

\begin{figure*}[htbp]
  \centering
  \subfigure[Service Degradation Obstructing Service Resilience Goal]{
      \centering
      \includegraphics[width=.4\linewidth]{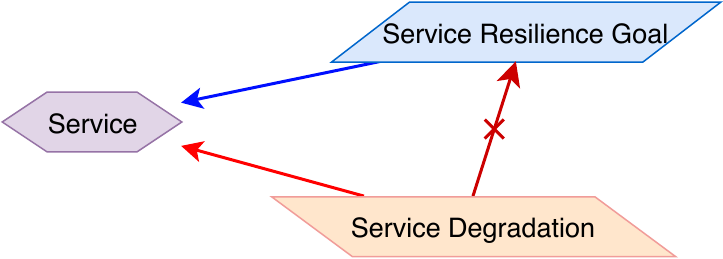}
  }
  \subfigure[Disruption Obstructing Resource Resilience Goal]{
      \centering
      \includegraphics[width=.44\linewidth]{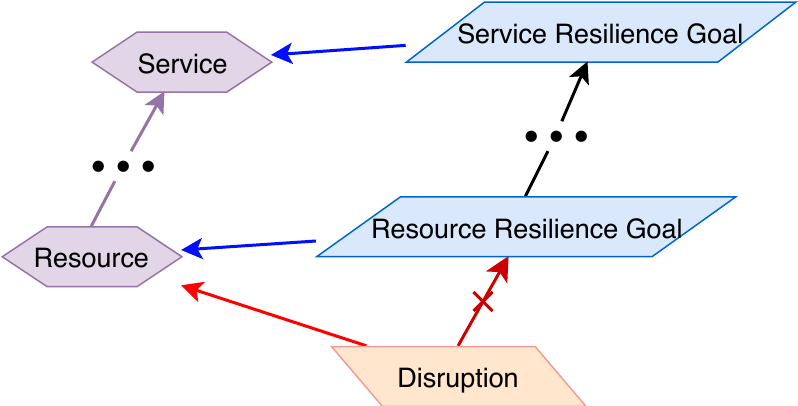}
  }
  \caption{A Service Degradation(a) and its Root Cause Disruption(b)}
  \label{fig:degrad-disrupt}
\end{figure*}

\textbf{Obstacle} is a dual notion to goal in goal models, 
and is represented by red parallelograms.
An obstacle links to goals with the \textbf{obstructs} link to show the goals can't be satisfied when the obstacle gets true, and is linked by a goal representing the corresponding resilience mechanism with the \textbf{resolve} link, as is shown in Figure \ref{fig:total-refinement}.
In our goal model, service degradations are obstacles to service resilience goals.
And a service degradation is transformed into a disruption obstructing a resource resilience goal after the service degradation is diagnosed, as is shown in Figure \ref{fig:degrad-disrupt}.
The disruption is linked to the affected system resource with the \textbf{affects} link, and then corresponding resilience mechanism will take actions to the affected resource.

The textual specification of an obstacle contains the following information:
\begin{itemize}
\bitem{Obstacle Name} The identifier of the obstacle;
\bitem{Event} The description that how service/resource resilience goals are obstructed.
\end{itemize}

\bigskip

\subsubsection{Asset}
\

When considering resilience of MSA systems, services and system resources are victims of disruptions and entities to be manipulated to recover from service degradations.
\textbf{Assets} are used represent services or system resources(like containers, physical servers) in an MSA System.
Assets are represented by purple hexagons in diagrams of our proposed goal model.
Textual specification of an asset includes the asset name and the asset type.

An asset is linked to another asset with the \textbf{support} link, to show that a system resource supports the running of a service/another system resource resource. 
For example, a service asset is linked with assets representing service's dependency system resources.
The support link will be used as references to service resilience goal refinement and service degradation diagnosis.

\bigskip

\subsubsection{System Components}

In MSA Systems, System components like anomaly detectors, monitoring tools are required to execute system behaviors decomposed from resilience mechanisms.
In our goal model, we used \textbf{agents} in KAOS to represent these components.
System components are shown as yellow hexagons, and they link to the system behaviors that they are expected to perform with the \textbf{executes} link.
Same with assets, textual information of system components includes the agent name and the agent type.

\bigskip

\subsubsection{Domain Property}

\textbf{Domain properties} are indicative statements of domain knowledge which are used as references of elements in KAOS\cite{van2001goal}\cite{lamsweerde2003kaos}. 
Service performance benchmarks are domain properties of service resilience goals, 
because service performance benchmarks are key variables to calculate service resilience metrics in MRMM.
Domain knowledges of MSA Systems (like architectural patterns, operation principles) are domain properties of resilience mechanisms, 
because the selection of a resilience mechanism depends on what microservice component or technology stack is used \cite{garriga2017towards}.
A domain property is represented by an orange pentagon, and linked with a goal it supplements with the \textbf{refers to} link,
and textual specifications of domain properties include the following information:
\begin{itemize}
\bitem{Domain Property Name} The identifier of the domain property;
\bitem{Description} Detailed description of the domain property;
\bitem{Reference Resources} Reference links to related documentations of the domain property.

\end{itemize}

\subsection{Resilience Mechanism Implementation View}

Resilience Mechanism Implementation View shows how a resilience mechanism, which is established in Resilience Goal Decomposition View, is implemented in an MSA System.
Existing documentation styles for resilience mechanisms (like architecture-level diagrams used for Circuit Breakers and Bulkheads in \cite{newman2015building}\cite{nygard2018release}) are directly used for Resilience Mechanism Implementation View, 
because there is no need to design a new model since these documentation styles are expressive enough.

Resilience Mechanism Implementation View is drawn according to the elements in Resilience Goal Decomposition View.
For example, if the component-and-connector architecture style\cite{clements2002documenting} is used for Resilience Mechanism Implementation View, each agent in Resilience Goal Decomposition View is drawn as boxes representing components, and system behaviors of resilience mechanisms are drawn as connections launched by these components.

\section{Case Study}
\label{sec:case}

In order to verify the feasibility of MRMM and Service Resilience Requirement Model, we conducted a case study on an MSA System.
One of the benchmark MSA Systems proposed in the literature\cite{Aderaldo2017Benchmark} was used as our target system.
We developed an open source web application called \textit{KAOSer}
\footnote{\url{https://github.com/XLab-Tongji/KAOSer}} was to draw diagrams of Service Resilience Requirement Model.

\begin{figure}
\centering
\includegraphics[width=.6\linewidth]{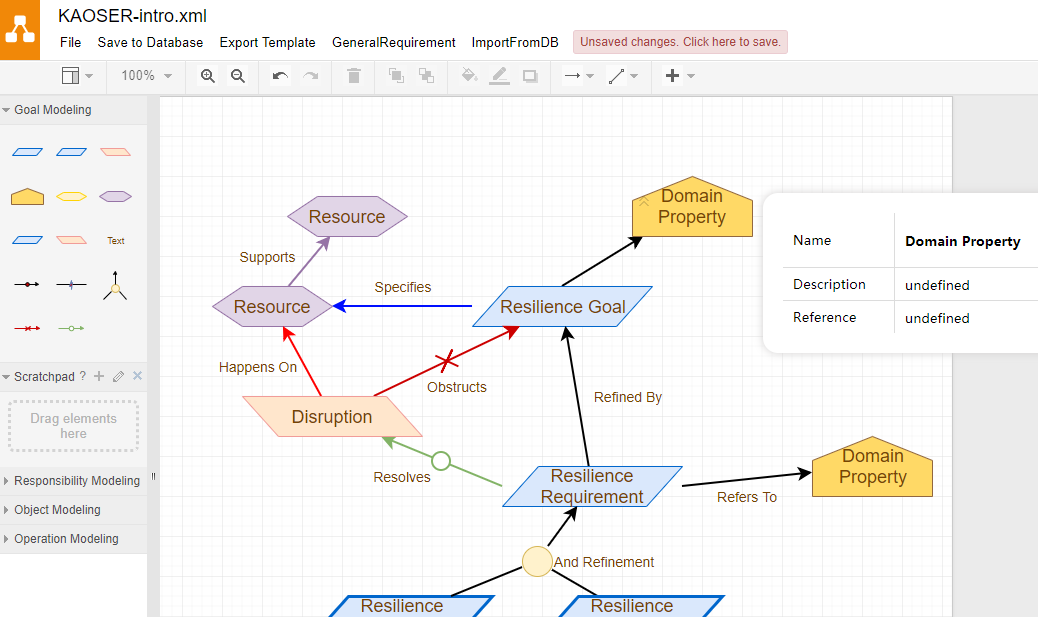}
\caption{User Interface of KAOSer}
\label{fig:kaoser}
\end{figure}

In order to draw diagrams of Resilience Goal Decomposition View, we developed an open source web application called \textit{KAOSer}
\footnote{\url{https://github.com/XLab-Tongji/KAOSer}}.
Users can create, edit elements in Resilience Goal Decomposition View, and export the diagram and textual documentation to their local storage.
Figure \ref{fig:kaoser} shows the user interface of \textit{KAOSer}.
Users can drag elements to the diagram area from the toolbox on the left,
and edit the textual specifications of these elements by clicking on these elements.

\subsection{System Description}
\label{sec:case-sys}
\begin{figure*}[htbp]
\centering
\subfigure[Architecture]{
  \includegraphics[width=.44\linewidth]{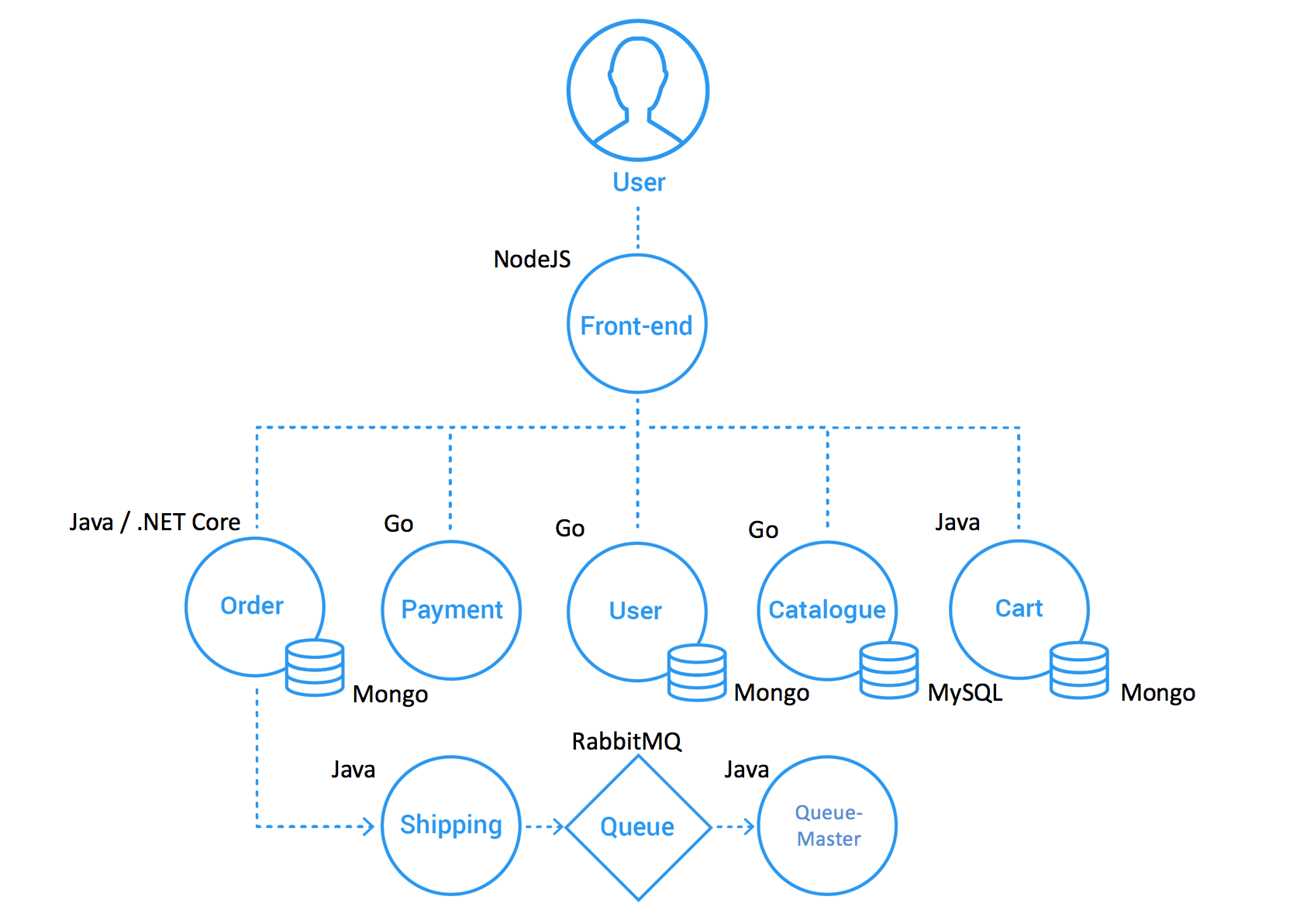}
  \label{fig:sock-archi}
}
\subfigure[Deployment Scenario]{
  \includegraphics[width=.44\linewidth]{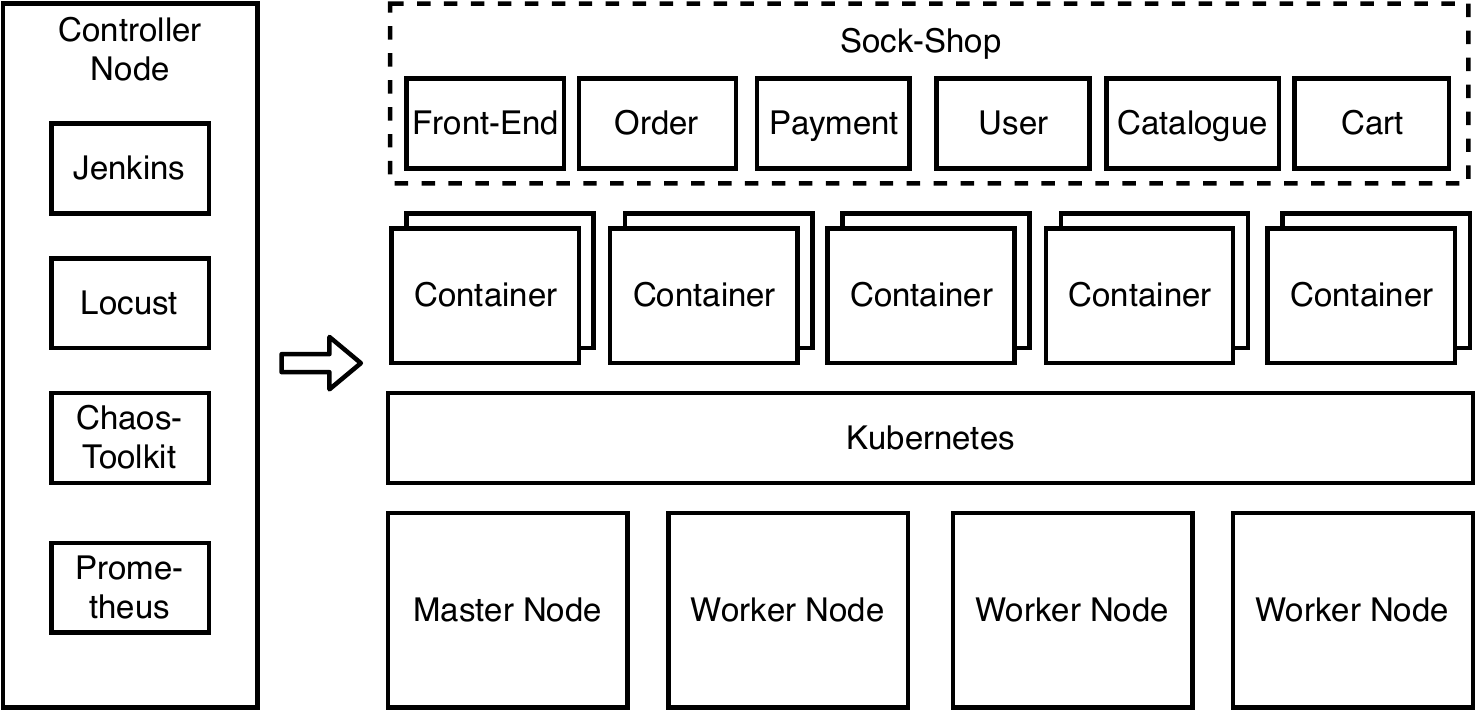}
  \label{fig:sock-deploy}
}
\caption{Target System: Sock Shop}
\label{fig:target-system}
\end{figure*}

Sock Shop \footnote{\url{https://github.com/microservices-demo/microservices-demo}} is an open source MSA System for demonstration and testing of microservice and cloud native technologies.
It is built using \textit{Spring Boot}, \textit{Go kit} and \textit{Node.js},
and is packaged in \textit{Docker} containers.
Sock Shop provides basic services of an online shopping system, Figure \ref{fig:sock-archi} shows the architecture of these services.

We deployed Sock Shop on a \textit{Kubernetes} cluster with one master node and three worker nodes. 
A Controller Node (which contains a bunch of tools including deployment, workload simulation, performance monitoring, fault injection, etc.) was used to generate and collect necessary data for resilience requirement elicitation. Figure \ref{fig:sock-deploy} shows the deployment scenario of Sock Shop in our case study.

\subsection{Resilience Requirement Representation}

In our proposed resilience requirement model, service resilience goals are set first.
Then service resilience goals are refined to resource resilience goals, disruptions violating service/resource resilience goals are identified and resilience mechanisms are established.
Limited by the length of the paper, we cannot list all service resilience goals, disruptions and resilience mechanism in our case study.
So we just illustrate a typical example in each step.

\subsubsection{Service Resilience Goal Setting}
Service Resilience Goal are thresholds of resilience metrics in MRMM. 
According to our definitions in MRMM, resilience metrics are derived from performance benchmarks of services.
So services' performance attributes and corresponding performance benchmarks are determined first.

Considering the metrics collectable from the monitoring tool \textit{Prometheus}, and existing common researches \& standards on web service qualities,
we chose \textit{response time} and \textit{success rate} as the general service performance attributes because all services in Sock Shop are transactional services.
Since the final goal of resilience is to prevent business loss caused by service degradations, we also set some performance attributes on business for certain services besides general service performance attributes.
We used the number of success orders, add-to-cart operations and online users per second, as our business performance attributes, because they are collectable and directly impact the "revenue" of Shop Shop.
Table \ref{tab:sock-qos} shows all service performance attributes selected for Sock Shop. 

\begin{table}
\centering
\caption{Service Performance Attributes of Sock Shop}
\label{tab:sock-qos}
\begin{tabular}{m{5em} m{4.5em} m{12em} m{3em}}
\toprule
Service & Performance Attribute & Description & Unit \\
\tablespace\colrule
All Services		&
Response Time		& 	Average Time taken to send a request and receive a response		& ms \\
\tablespace
&Success Rate	&	Number of response / number of request messages			& \%		\\
\tablespace
Order	&	
Success Orders		& 	Number of finished order per second						& orders/s	\\
\tablespace
Cart				&
Add-Cart Count		& 	Number of add-to-cart requests per second				& request/s	\\
\tablespace
User				&
Online Users		&	Number of online users									& users		\\
\tablespace\botrule

\end{tabular}
\end{table}

\begin{figure}
\centering
\includegraphics[width=\linewidth]{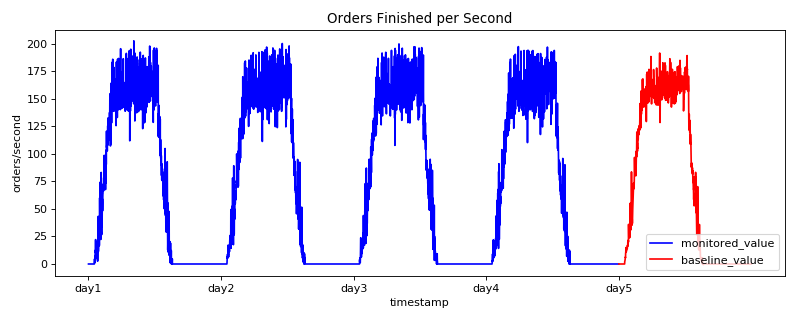}
\caption{Success Orders of Sock Shop under the Simulated Workload(the Blue Line), and the Performance Benchmark(the Red Line)}
\label{fig:order-perf}
\end{figure}

By referring to mean values of running data we simulated, 
we set \textbf{3 seconds} and \textbf{90\%} as the benchmark value of Response Time and Success Rate.
For performance attributes varying from time like Success Orders, we used time series prediction algorithms to build performance benchmarks with the workload data we simulated, as is shown in Figure \ref{fig:order-perf}.

For resilience goal setting, we only show the example of the Order service here.
The Order service has three performance attributes: Response Time, Success Rate and Success Orders.
Not all thershold values of three resilience metrics are set because some of them are meaningless(e.g. Performance Loss of Response Time).
We set thresholds of Disruption Tolerance and Recovery Time by referring to suggested values in ETSI standards.
For Performance Loss, by assuming that 5\% loss of orders in a day is not expected in Sock Shop, the Performance Loss thresholds for Success Orders of the Order service was roughly calculated.
Table \ref{tab:sock-sg} shows thresholds in resilience goals of the Order service.

\begin{table}
\centering
\caption{Service Resilience Goal of the Order Service}
\label{tab:sock-sg}
\begin{tabular}{m{10em} m{4em} m{4em} m{4em}}
\toprule
Performance Attribute & Disruption Tolerance & Recovery Time & Performance Loss \\
\colrule
Response Time	& 10s	& 5s	& - \\
\tablespace
Success Rate	& 20\%	& -		& -	\\
\tablespace
Success Orders	& -		& -		& 500 orders \\
\botrule
\end{tabular}
\end{table}

\subsubsection{Resource Resilience Goal Setting}
In this step, service resilience goals are refined to resource resilience goals.
Here we just show how the resilience goal of the Order service was decomposed.
The Order service has three performance attributes: Response Time, Success Rate and Success Orders.
In the Kubernetes cluster where Sock Shop was deployed,
the Order service was deployed on a pod which consists of several container instances.
Metrics of the Order service's performance attributes can be calculated by performance attributes of containers by the following formulas:

\begin{multline}
\label{eq:study-de-rt}
ResponseTime(Service) = \\
\frac{\sum_i^n{(ResponseTime(C_i)\times TPS(C_i))}}{\sum_i^n{TPS(C_i)}}
\end{multline}

\begin{multline}
\label{eq:study-de-sr}
SuccessRate(Service) = \\
\frac{\sum_i^n{(SuccessRate(C_i)\times TPS(C_i))}}{\sum_i^n{TPS(C_i)}}
\end{multline}

\begin{multline}
\label{eq:study-de-so}
SuccessOrders(Service) = \\
\sum_i^n{(SuccessRate(C_i)\times TPS(C_i))}
\end{multline}

In Formula \ref{eq:study-de-rt}, \ref{eq:study-de-sr} and \ref{eq:study-de-so}, $n$ means the number of available container instances of a POD and $C_i$ means the $i$th container instance.
From these three formulas it can be found that performance attributes of the Order service can be calculated by performance attributes of its pods and containers.
So service resilience goals of the Order service were refined to resource resilience goals of the Order services' pod and containers.
TPS, Response Time, Success Rate of containers, and the number of available instances of a pod, were selected as resource performance attributes to be benchmarked.

We set the baseline value of Response Time and Success Rate of the Order container same with the performance benchmark of the Order service. 
Benchmark of the Order container's TPS is a dynamic value calculated by time series algorithms.
The baseline value of Available Instances in a pod was set to \textbf{3 instances}, according to the deployment environment of the target system. 

\begin{figure*}
\centering
\includegraphics[width=.7\linewidth]{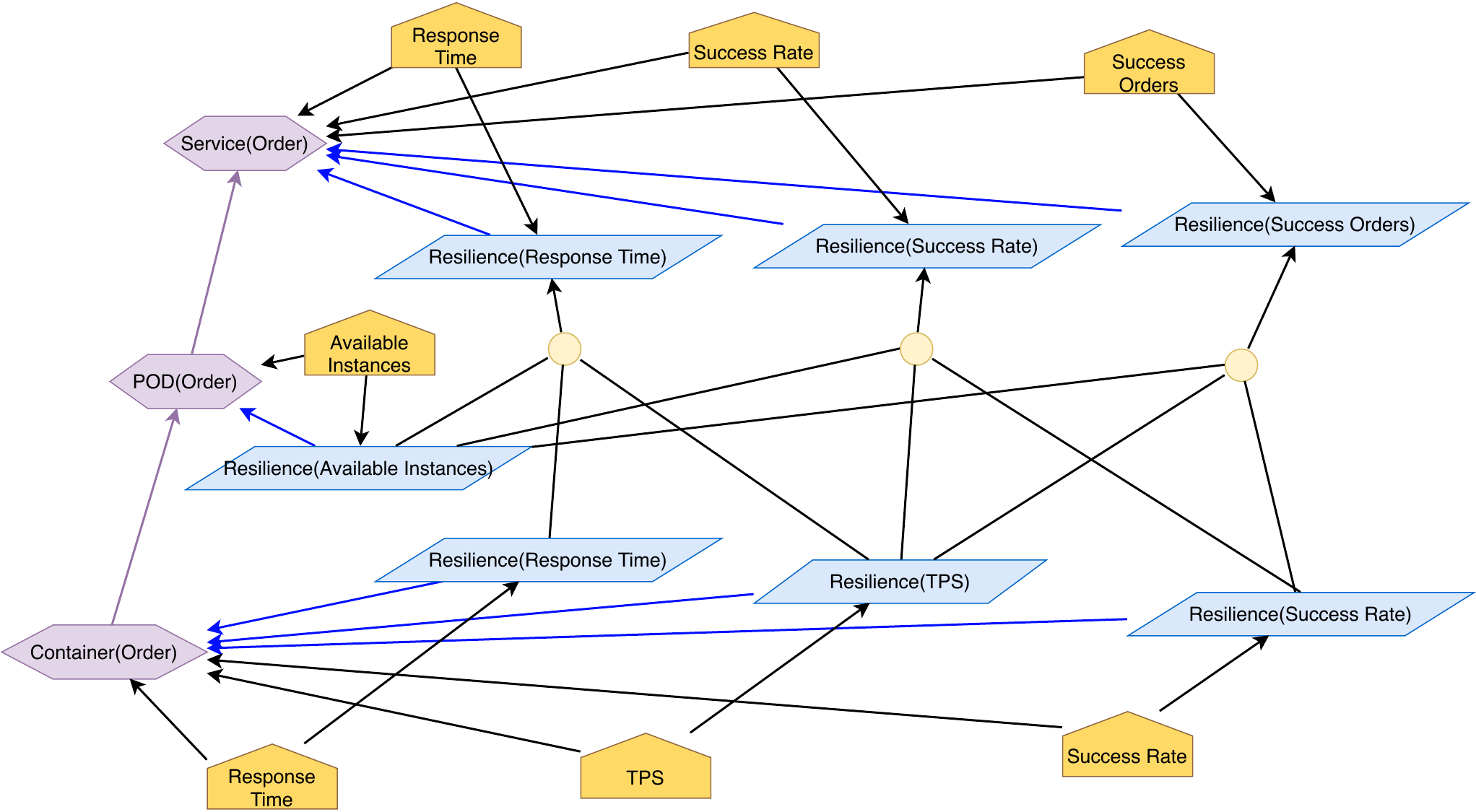}
\caption{Service Resilience Goal Refinement for the Order service}
\label{fig:sock-order}
\end{figure*}

The Order container's thresholds of Disruption Tolerance and Recovery Time were the same with thresholds of the order service.
Threshold value of TPS's Performance Loss was calculated by the resilience goal of the Order service.
For Available Instance of the Order pod, we assumed that there is at most one instance will break down and it will be recovered within 2 seconds.
Table \ref{tab:sock-rg} shows thresholds in resource resilience goals refined from the Order service's resilience goal. Figure \ref{fig:sock-order} shows the goal refinement of the Order service in Service Resilience Requirement Model.

\begin{table}
\centering
\caption{Resource Resilience Goals Refined from the Order Service's Resilience Goal}
\label{tab:sock-rg}
\begin{tabular}{m{7em} m{5em} m{4em} m{4em} m{4em}}
\toprule
Resource & Performance Attribute & Disruption Tolerance & Recovery Time & Performance Loss \\
\colrule
Order(POD)		&	Available Instances &	1 instance	&	2s	&	-	\\
\colrule
Order(Container)& 	Response Time	& 10s	& 5s	& - \\
\tablespace
				&	Successability	& 20\%	& -		& -	\\
\tablespace
				& 	TPS				&	-	&	-	&	150	transactions	\\
\botrule
\end{tabular}
\end{table}

\subsubsection{Resilience Mechanism Establishment}

\begin{figure}
\centering
\includegraphics[width=\linewidth]{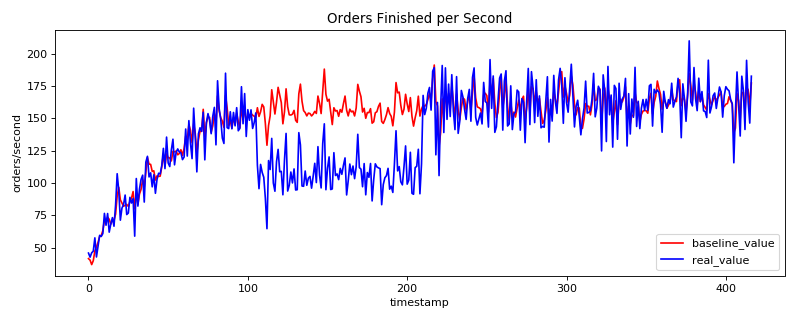}
\caption{A Service Degradation Violating Service Resilience Goal of the Order Service}
\label{fig:sock-degrad}
\end{figure}

\begin{figure}
\centering
\includegraphics[width=.8\linewidth]{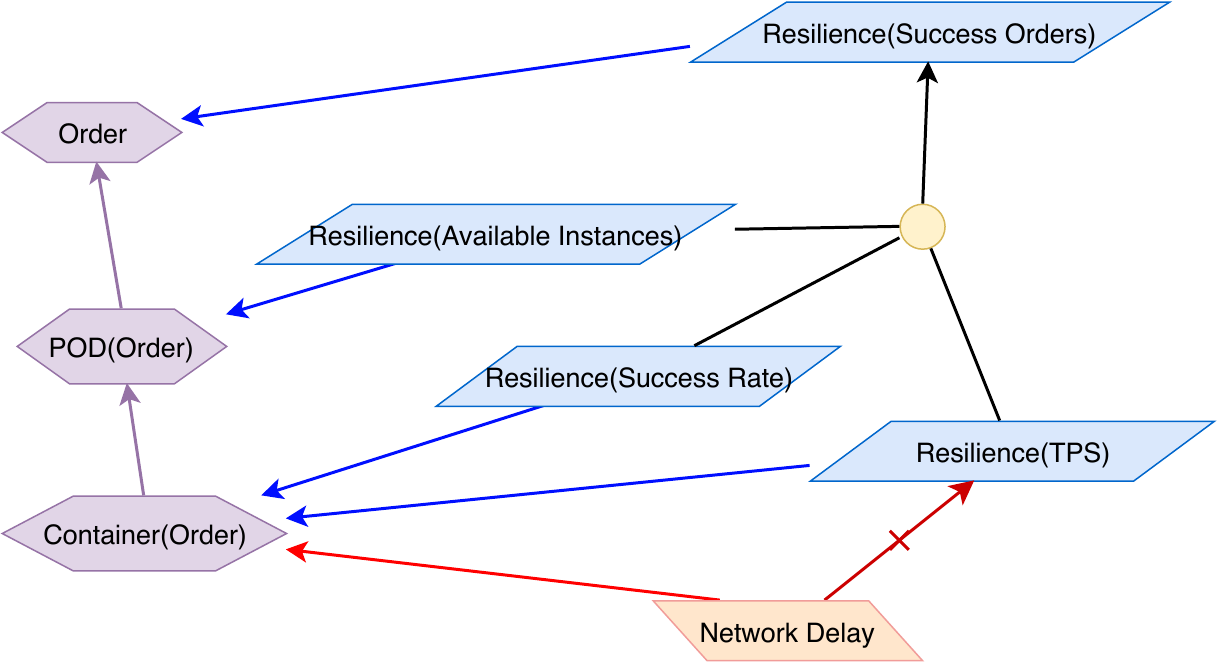}
\caption{Root Cause Disruption of the Service Degradation in Figure \ref{fig:sock-degrad}}
\label{fig:sock-dis}
\end{figure}

Reslience mechanisms are established to solve disruptions that violate resilience goals.
In this case study, we proactively injected random faults to our 24*7 running Sock Shop to detect service degradations violating service resilience goals we set. Figure \ref{fig:sock-degrad} shows a service degradation detected on the Order Service that violated the service resilience goal on Success Orders.
By searching fault injection logs and related monitoring data, we found that the service degradation was caused by the network delay we injected to containers of the Order service, and it also violated resilience goals on TPS of the Order service's container, as is shown in Figure \ref{fig:sock-dis}.

\begin{figure}
\centering
\includegraphics[width=.8\linewidth]{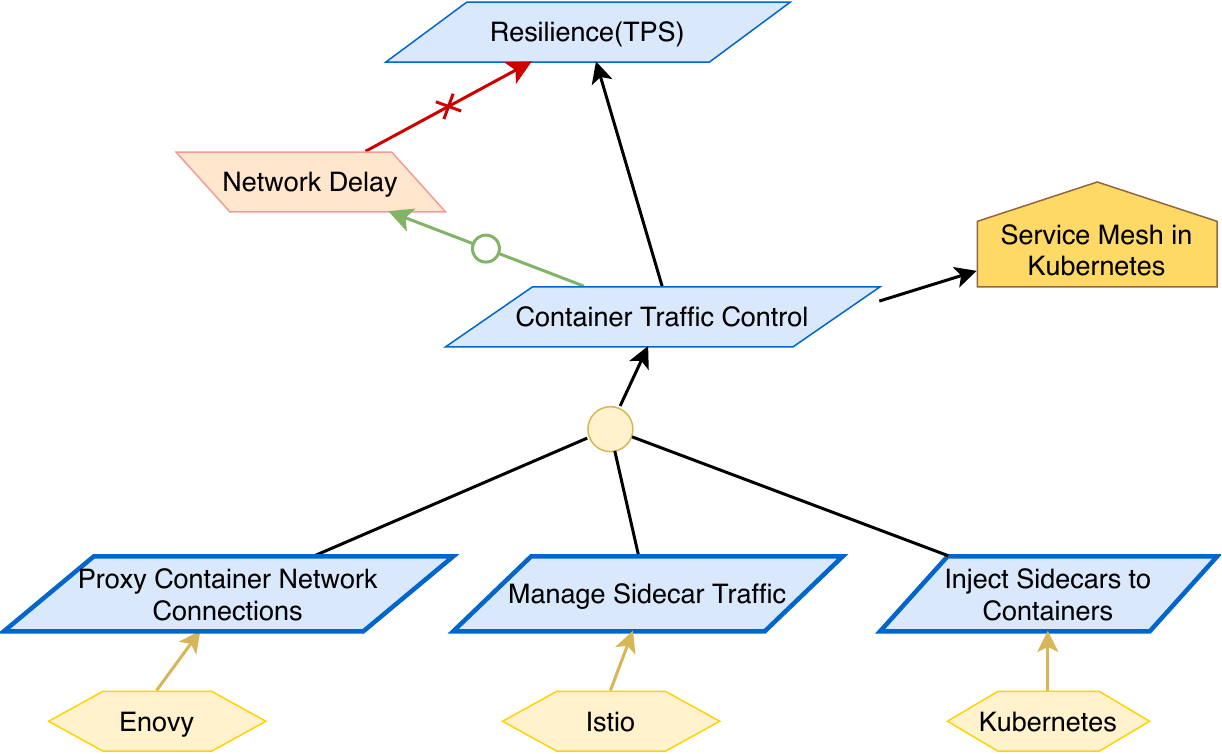}
\caption{Resilience Mechanism Established for Container Network Delay}
\label{fig:dis-resolv}
\end{figure}

\begin{figure}
\centering
\includegraphics[width=.8\linewidth]{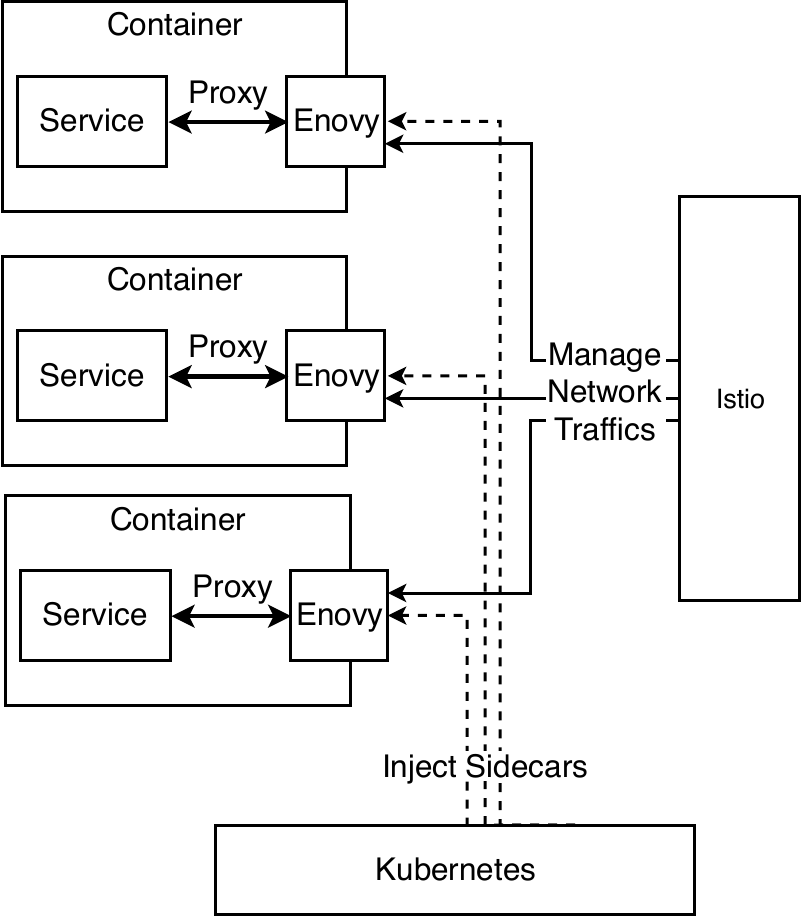}
\caption{Resilience Mechanism in Resilience Mechanism Implementation View}
\label{fig:svc-mesh}
\end{figure}

1
Network delays on containers impact the container's ability to process transactions,
so we planned to transfer more transactions to other normal container instances when network delay on a container is detected.
Since the target system is deployed on a Kubernetes cluster, this resilience mechanism can use \textit{Service Mesh} to control network traffics among containers.
In order to achieve it, three jobs are required: 
\begin{enumerate}
\item The Kubernetes cluster injects the sidecar component \textit{Enovy} into all containers deployed on the cluster;
\item Enovy proxies network connections of all containers, and transmits network packages to the Service Mesh middleware \textit{Istio};
\item Istio monitors and manages network traffic among sidecars.
\end{enumerate} 
Figure \ref{fig:dis-resolv} shows the resilience mechanism and system behaviors refined in Resilience Goal Decomposition View, 
and Figure \ref{fig:svc-mesh} shows the corresponding representation in Resilience Mechanism Implementation View. 

\section{Conclusion}
\label{sec:conclusion}

In recent years, the microservice architecture has already been a mainstream architecture adopted by internet companies.
In cases when achieving higher system reliability is no longer affordable and failure is inevitable \cite{beyer2016site}, microservice practitioners start to use the word "resilience" to describe the ability coping with failures.
However, due to no consensus on definitions and measurements of resilience, microservice practitioners seldom have a clear idea on how to evaluate the resilience of an MSA System and how resilient an MSA System is supposed to be. 

In this work, we have the following contributions:
\begin{itemize}
\item The definition of microservice resilience is provided by referring to systematic studies on resilience in other scientific areas. And a Microservice Reslience Measurement Model is proposed to measure service resilience of service degradations. 
\item Service Resilience Requirement Model is proposed to represent service resilience goals, disruptions and resilience mechanisms.
The requirement model uses goal model to refine service resilience goals with thresholds of resilience metrics in MRMM, to system behaviors to be implemented in MSA Systems.
\end{itemize}

Possible limitation of our proposed work, which can be improved in future work, include the followings:
\begin{itemize}
\item The MRMM model gives measurable resilience metrics to evaluate a service's resilience,
and it may be worthwhile to find a modeling technique to model system resilience with service resilience.
\item Verification on goal satisfaction is an important type of research in Goal Modeling.
We have already given mathematical presentations of notions in our measurement model and requirement model.
How to encode these mathematical presentations into formal languages, and verify them with model checkers, are future works of this paper.
\item In our case study, an open source microservice project was used as the target system. The target system contains not too many microservice so that we can easily build resilience documents with manual efforts. 
For large scale MSA Systems, generating resilience requirements completely manually is difficult because there are a lot of services and resources to be identified. 
So auto generation techniques (e.g. Generate the architecture of the system by reading deployment configuration files) for our resilience requirement model is needed.
\end{itemize}

\section{Acknowledgement}
The paper is partially supported by Key Technologies of Integrated Intelligent Construction of “Collection-Design-Construction” for High-speed Rail Tunnel Fund of NSFC (No.U1934212).

\bibliographystyle{unsrt}
\bibliography{ref} 

\begin{thebibliography}{10}

\bibitem{fowler2014microservices}
Martin Fowler and James Lewis.
\newblock Microservices.
\newblock {\em ThoughtWorks. http://martinfowler. com/articles/microservices.
  html [last accessed on February 17, 2015]}, 2014.

\bibitem{hoff2017lessons}
Todd Hoff.
\newblock Lessons learned from scaling uber to 2000 engineers, 1000 services,
  and 8000 git repositories, 2017.

\bibitem{mauro2016adopting}
TONY Mauro.
\newblock Adopting microservices at netflix: Lessons for architectural design,
  2016.

\bibitem{newman2015building}
Sam Newman.
\newblock {\em Building Microservices}.
\newblock " O'Reilly Media, Inc.", 2015.

\bibitem{humble2010continuous}
Jez Humble and David Farley.
\newblock {\em Continuous Delivery: Reliable Software Releases through Build,
  Test, and Deployment Automation (Adobe Reader)}.
\newblock Pearson Education, 2010.

\bibitem{bass2015devops}
Len Bass, Ingo Weber, and Liming Zhu.
\newblock {\em DevOps: A Software Architect's Perspective}.
\newblock Addison-Wesley Professional, 2015.

\bibitem{balalaie2016microservices}
Armin Balalaie, Abbas Heydarnoori, and Pooyan Jamshidi.
\newblock Microservices architecture enables devops: migration to a
  cloud-native architecture.
\newblock {\em IEEE Software}, 33(3):42--52, 2016.

\bibitem{hatton1997reexamining}
Les Hatton.
\newblock Reexamining the fault density component size connection.
\newblock {\em IEEE software}, 14(2):89--97, 1997.

\bibitem{montesi2016circuit}
Fabrizio Montesi and Janine Weber.
\newblock Circuit breakers, discovery, and api gateways in microservices.
\newblock {\em arXiv preprint arXiv:1609.05830}, 2016.

\bibitem{esposito2016challenges}
Christian Esposito, Aniello Castiglione, and Kim-Kwang~Raymond Choo.
\newblock Challenges in delivering software in the cloud as microservices.
\newblock {\em IEEE Cloud Computing}, 3(5):10--14, 2016.

\bibitem{Gunawi2016Why}
Haryadi~S. Gunawi, Mingzhe Hao, Riza~O. Suminto, Agung Laksono, and Kurnia~J.
  Eliazar.
\newblock Why does the cloud stop computing?: Lessons from hundreds of service
  outages.
\newblock In {\em Acm Symposium on Cloud Computing}, 2016.

\bibitem{organizacion2011iso}
Organizaci{\'o}n~Internacional de~Normalizaci{\'o}n.
\newblock {\em ISO-IEC 25010: 2011 Systems and Software Engineering-Systems and
  Software Quality Requirements and Evaluation (SQuaRE)-System and Software
  Quality Models}.
\newblock ISO, 2011.

\bibitem{Gunawi2014What}
Haryadi~S. Gunawi, Mingzhe Hao, Tanakorn Leesatapornwongsa, Tiratat
  Patana-Anake, Thanh Do, Jeffry Adityatama, Kurnia~J. Eliazar, Agung Laksono,
  Jeffrey~F. Lukman, and Vincentius Martin.
\newblock What bugs live in the cloud? a study of 3000+ issues in cloud
  systems.
\newblock 2014.

\bibitem{gunawi2018fail}
Haryadi~S Gunawi, Riza~O Suminto, Russell Sears, Casey Golliher, Swaminathan
  Sundararaman, Xing Lin, Tim Emami, Weiguang Sheng, Nematollah Bidokhti,
  Caitie McCaffrey, et~al.
\newblock Fail-slow at scale: Evidence of hardware performance faults in large
  production systems.
\newblock {\em ACM Transactions on Storage (TOS)}, 14(3):23, 2018.

\bibitem{wolff2016microservices}
Eberhard Wolff.
\newblock {\em Microservices: Flexible Software Architecture}.
\newblock Addison-Wesley Professional, 2016.

\bibitem{nadareishvili2016microservice}
Irakli Nadareishvili, Ronnie Mitra, Matt McLarty, and Mike Amundsen.
\newblock {\em Microservice Architecture: Aligning Principles, Practices, and
  Culture}.
\newblock " O'Reilly Media, Inc.", 2016.

\bibitem{bondavalli2009research}
A~Bondavalli et~al.
\newblock Research roadmap deliverable d3. 2.
\newblock {\em AMBER--Assessing Measuring and Benchmarking Resilience, Funded
  by European Union}, 2009.

\bibitem{almeida2012changeloads}
Raquel Almeida and Marco Vieira.
\newblock Changeloads for resilience benchmarking of self-adaptive systems: a
  risk-based approach.
\newblock In {\em Dependable Computing Conference (EDCC), 2012 Ninth European},
  pages 173--184. IEEE, 2012.

\bibitem{evans2004domain}
Eric Evans.
\newblock {\em Domain-driven design: tackling complexity in the heart of
  software}.
\newblock Addison-Wesley Professional, 2004.

\bibitem{DBLP:journals/software/RademacherSS18}
Florian Rademacher, Jonas Sorgalla, and Sabine Sachweh.
\newblock Challenges of domain-driven microservice design: {A} model-driven
  perspective.
\newblock {\em {IEEE} Softw.}, 35(3):36--43, 2018.

\bibitem{holling1973resilience}
Crawford~S Holling.
\newblock Resilience and stability of ecological systems.
\newblock {\em Annual review of ecology and systematics}, 4(1):1--23, 1973.

\bibitem{righi2015systematic}
Angela~Weber Righi, Tarcisio~Abreu Saurin, and Priscila Wachs.
\newblock A systematic literature review of resilience engineering: Research
  areas and a research agenda proposal.
\newblock {\em Reliability Engineering \& System Safety}, 141:142--152, 2015.

\bibitem{hosseini2016review}
Seyedmohsen Hosseini, Kash Barker, and Jose~E Ramirez-Marquez.
\newblock A review of definitions and measures of system resilience.
\newblock {\em Reliability Engineering \& System Safety}, 145:47--61, 2016.

\bibitem{ungar2003qualitative}
Michael Ungar.
\newblock Qualitative contributions to resilience research.
\newblock {\em Qualitative social work}, 2(1):85--102, 2003.

\bibitem{bruneau2003framework}
Michel Bruneau, Stephanie~E Chang, Ronald~T Eguchi, George~C Lee, Thomas~D
  O’Rourke, Andrei~M Reinhorn, Masanobu Shinozuka, Kathleen Tierney,
  William~A Wallace, and Detlof Von~Winterfeldt.
\newblock A framework to quantitatively assess and enhance the seismic
  resilience of communities.
\newblock {\em Earthquake spectra}, 19(4):733--752, 2003.

\bibitem{henry2012generic}
Devanandham Henry and Jose~Emmanuel Ramirez-Marquez.
\newblock Generic metrics and quantitative approaches for system resilience as
  a function of time.
\newblock {\em Reliability Engineering \& System Safety}, 99:114--122, 2012.

\bibitem{yodo2016engineering}
Nita Yodo and Pingfeng Wang.
\newblock Engineering resilience quantification and system design implications:
  A literature survey.
\newblock {\em Journal of Mechanical Design}, 138(11):111408, 2016.

\bibitem{dragoni2017microservices}
Nicola Dragoni, Saverio Giallorenzo, Alberto~Lluch Lafuente, Manuel Mazzara,
  Fabrizio Montesi, Ruslan Mustafin, and Larisa Safina.
\newblock Microservices: yesterday, today, and tomorrow.
\newblock In {\em Present and Ulterior Software Engineering}, pages 195--216.
  Springer, 2017.

\bibitem{dragoni2017microserviceshow}
Nicola Dragoni, Ivan Lanese, Stephan~Thordal Larsen, Manuel Mazzara, Ruslan
  Mustafin, and Larisa Safina.
\newblock Microservices: How to make your application scale.
\newblock {\em arXiv preprint arXiv:1702.07149}, 2017.

\bibitem{Brilhante:2017:AQB:3126858.3126873}
Jonathan Brilhante, Rostand Costa, and Tiago Maritan.
\newblock Asynchronous queue based approach for building reactive
  microservices.
\newblock In {\em Proceedings of the 23rd Brazillian Symposium on Multimedia
  and the Web}, WebMedia '17, pages 373--380, New York, NY, USA, 2017. ACM.

\bibitem{7965189}
J.~{Jenkins}, G.~{Shipman}, J.~{Mohd-Yusof}, K.~{Barros}, P.~{Carns}, and
  R.~{Ross}.
\newblock A case study in computational caching microservices for hpc.
\newblock In {\em 2017 IEEE International Parallel and Distributed Processing
  Symposium Workshops (IPDPSW)}, pages 1309--1316, May 2017.

\bibitem{8486300}
Y.~{Niu}, F.~{Liu}, and Z.~{Li}.
\newblock Load balancing across microservices.
\newblock In {\em IEEE INFOCOM 2018 - IEEE Conference on Computer
  Communications}, pages 198--206, April 2018.

\bibitem{8530769}
M.~{Cinque}, R.~{Della Corte}, R.~{Iorio}, and A.~{Pecchia}.
\newblock An exploratory study on zeroconf monitoring of microservices systems.
\newblock In {\em 2018 14th European Dependable Computing Conference (EDCC)},
  pages 112--115, Sep. 2018.

\bibitem{Montesi:2018:DPC:3167132.3167427}
Fabrizio Montesi and Janine Weber.
\newblock From the decorator pattern to circuit breakers in microservices.
\newblock In {\em Proceedings of the 33rd Annual ACM Symposium on Applied
  Computing}, SAC '18, pages 1733--1735, New York, NY, USA, 2018. ACM.

\bibitem{richter2017highly}
Daniel Richter, Marcus Konrad, Katharina Utecht, and Andreas Polze.
\newblock Highly-available applications on unreliable infrastructure:
  Microservice architectures in practice.
\newblock In {\em Software Quality, Reliability and Security Companion (QRS-C),
  2017 IEEE International Conference on}, pages 130--137. IEEE, 2017.

\bibitem{kratzke2017microservices}
Nane Kratzke.
\newblock About microservices, containers and their underestimated impact on
  network performance.
\newblock {\em arXiv preprint arXiv:1710.04049}, 2017.

\bibitem{toffetti2015architecture}
Giovanni Toffetti, Sandro Brunner, Martin Bl{\"o}chlinger, Florian Dudouet, and
  Andrew Edmonds.
\newblock An architecture for self-managing microservices.
\newblock In {\em Proceedings of the 1st International Workshop on Automated
  Incident Management in Cloud}, pages 19--24. ACM, 2015.

\bibitem{soenen2017optimising}
Thomas Soenen, Wouter Tavernier, Didier Colle, and Mario Pickavet.
\newblock Optimising microservice-based reliable nfv management \&
  orchestration architectures.
\newblock In {\em Resilient Networks Design and Modeling (RNDM), 2017 9th
  International Workshop on}, pages 1--7. IEEE, 2017.

\bibitem{zwietasch2017online}
Tim Zwietasch.
\newblock Online failure prediction for microservice architectures.
\newblock Master's thesis, 2017.

\bibitem{haselbock2017decision}
Stefan Haselb{\"o}ck, Rainer Weinreich, and Georg Buchgeher.
\newblock Decision guidance models for microservices: service discovery and
  fault tolerance.
\newblock In {\em Proceedings of the Fifth European Conference on the
  Engineering of Computer-Based Systems}, page~4. ACM, 2017.

\bibitem{heorhiadi2016gremlin}
Victor Heorhiadi, Shriram Rajagopalan, Hani Jamjoom, Michael~K Reiter, and Vyas
  Sekar.
\newblock Gremlin: systematic resilience testing of microservices.
\newblock In {\em Distributed Computing Systems (ICDCS), 2016 IEEE 36th
  International Conference on}, pages 57--66. IEEE, 2016.

\bibitem{brogi2017towards}
Antonio Brogi, Andrea Canciani, Davide Neri, Luca Rinaldi, and Jacopo Soldani.
\newblock Towards a reference dataset of microservice-based applications.
\newblock In {\em International Conference on Software Engineering and Formal
  Methods}, pages 219--229. Springer, 2017.

\bibitem{D2017Model}
Thomas~F. Düllmann and André~Van Hoorn.
\newblock Model-driven generation of microservice architectures for
  benchmarking performance and resilience engineering approaches.
\newblock In {\em The Acm/spec}, pages 171--172, 2017.

\bibitem{van2001goal}
Axel Van~Lamsweerde.
\newblock Goal-oriented requirements engineering: A guided tour.
\newblock In {\em Requirements Engineering, 2001. Proceedings. Fifth IEEE
  International Symposium on}, pages 249--262. IEEE, 2001.

\bibitem{rubin1992object}
Kenneth~S Rubin and Adele Goldberg.
\newblock Object behavior analysis.
\newblock {\em Communications of the ACM}, 35(9):48--62, 1992.

\bibitem{van2004object}
Axel Van~Lamsweerde and Emmanuel Letier.
\newblock From object orientation to goal orientation: A paradigm shift for
  requirements engineering.
\newblock In {\em Radical Innovations of Software and Systems Engineering in
  the Future}, pages 325--340. Springer, 2004.

\bibitem{dardenne1993goal}
Anne Dardenne, Axel Van~Lamsweerde, and Stephen Fickas.
\newblock Goal-directed requirements acquisition.
\newblock {\em Science of computer programming}, 20(1-2):3--50, 1993.

\bibitem{yu1997towards}
Eric~SK Yu.
\newblock Towards modelling and reasoning support for early-phase requirements
  engineering.
\newblock In {\em Requirements Engineering, 1997., Proceedings of the Third
  IEEE International Symposium on}, pages 226--235. IEEE, 1997.

\bibitem{DBLP:journals/aamas/BrescianiPGGM04}
Paolo Bresciani, Anna Perini, Paolo Giorgini, Fausto Giunchiglia, and John
  Mylopoulos.
\newblock Tropos: An agent-oriented software development methodology.
\newblock {\em Auton. Agents Multi Agent Syst.}, 8(3):203--236, 2004.

\bibitem{anton1996goal}
Annie~I Anton.
\newblock Goal-based requirements analysis.
\newblock In {\em Requirements Engineering, 1996., Proceedings of the Second
  International Conference on}, pages 136--144. IEEE, 1996.

\bibitem{mylopoulos1992representing}
John Mylopoulos, Lawrence Chung, and Brian Nixon.
\newblock Representing and using nonfunctional requirements: A process-oriented
  approach.
\newblock {\em IEEE Transactions on software engineering}, 18(6):483--497,
  1992.

\bibitem{lapouchnian2005goal}
Alexei Lapouchnian.
\newblock Goal-oriented requirements engineering: An overview of the current
  research.
\newblock {\em University of Toronto}, page~32, 2005.

\bibitem{rashid2002early}
Awais Rashid, Peter Sawyer, Ana Moreira, and Jo{\~a}o Ara{\'u}jo.
\newblock Early aspects: A model for aspect-oriented requirements engineering.
\newblock In {\em Requirements Engineering, 2002. Proceedings. IEEE Joint
  International Conference on}, pages 199--202. IEEE, 2002.

\bibitem{van1998integrating}
Axel Van~Lamsweerde and Emmanuel Letier.
\newblock Integrating obstacles in goal-driven requirements engineering.
\newblock In {\em Proceedings of the 20th international conference on Software
  engineering}, pages 53--62. IEEE Computer Society, 1998.

\bibitem{Van2000Handling}
Axel Van~Lamsweerde and Emmanuel Letier.
\newblock Handling obstacles in goal-oriented requirements engineering.
\newblock {\em IEEE Transactions on Software Engineering}, 26(10):978--1005,
  2000.

\bibitem{van2004elaborating}
Axel Van~Lamsweerde.
\newblock Elaborating security requirements by construction of intentional
  anti-models.
\newblock In {\em Proceedings of the 26th International Conference on Software
  Engineering}, pages 148--157. IEEE Computer Society, 2004.

\bibitem{wang2015discovering}
Hongbing Wang, Suxiang Zhou, and Qi~Yu.
\newblock Discovering web services to improve requirements decomposition.
\newblock In {\em Web Services (ICWS), 2015 IEEE International Conference on},
  pages 743--746. IEEE, 2015.

\bibitem{Zardari2011Cloud}
Shehnila Zardari and Rami Bahsoon.
\newblock Cloud adoption: a goal-oriented requirements engineering approach.
\newblock In {\em International Workshop on Software Engineering for Cloud
  Computing}, pages 29--35, 2011.

\bibitem{DBLP:journals/jss/ChungHLSDS13}
Lawrence Chung, Tom Hill, Owolabi Legunsen, Zhenzhou Sun, Adip Dsouza, and Sam
  Supakkul.
\newblock A goal-oriented simulation approach for obtaining good private
  cloud-based system architectures.
\newblock {\em J. Syst. Softw.}, 86(9):2242--2262, 2013.

\bibitem{DBLP:conf/sac/JuniorRSM15}
Ronaldo~Gon{\c{c}}alves Junior, Tiago Rolim, Am{\'{e}}rico Sampaio, and
  Nabor~C. Mendon{\c{c}}a.
\newblock A multi-criteria approach for assessing cloud deployment options
  based on non-functional requirements.
\newblock In Roger~L. Wainwright, Juan~Manuel Corchado, Alessio Bechini, and
  Jiman Hong, editors, {\em Proceedings of the 30th Annual {ACM} Symposium on
  Applied Computing, Salamanca, Spain, April 13-17, 2015}, pages 1383--1389.
  {ACM}, 2015.

\bibitem{patil2015cloud}
Shruti Patil and Roshani Ade.
\newblock Cloud data security for goal driven global software engineering
  projects.
\newblock {\em Procedia Computer Science}, 46:548--557, 2015.

\bibitem{deprez2012integrating}
Jean-Christophe Deprez, Ravi Ramdoyal, and Christophe Ponsard.
\newblock Integrating energy and eco-aware requirements engineering in the
  development of services-based applications on virtual clouds.
\newblock In {\em InFirst International Workshop on Requirements Engineering
  for Sustainable Systems}, 2012.

\bibitem{DBLP:journals/tse/DubocLR13}
Leticia Duboc, Emmanuel Letier, and David~S. Rosenblum.
\newblock Systematic elaboration of scalability requirements through
  goal-obstacle analysis.
\newblock {\em {IEEE} Trans. Software Eng.}, 39(1):119--140, 2013.

\bibitem{DBLP:conf/sac/ZardariBE14}
Shehnila Zardari, Rami Bahsoon, and Anik{\'{o}} Ek{\'{a}}rt.
\newblock Cloud adoption: prioritizing obstacles and obstacles resolution
  tactics using {AHP}.
\newblock In Yookun Cho, Sung~Y. Shin, Sang{-}Wook Kim, Chih{-}Cheng Hung, and
  Jiman Hong, editors, {\em Symposium on Applied Computing, {SAC} 2014,
  Gyeongju, Republic of Korea - March 24 - 28, 2014}, pages 1013--1020. {ACM},
  2014.

\bibitem{zardari2013using}
Shehnila Zardari, Funmilade Faniyi, and Rami Bahsoon.
\newblock Using obstacles for systematically modeling, analysing, and
  mitigating risks in cloud adoption.
\newblock In {\em Aligning Enterprise, System, and Software Architectures},
  pages 275--296. IGI Global, 2013.

\bibitem{Giorgini2002Reasoning}
Paolo Giorgini, John Mylopoulos, Eleonora Nicchiarelli, and Roberto Sebastiani.
\newblock {\em Reasoning with Goal Models}.
\newblock 2002.

\bibitem{clements2002documenting}
Paul Clements, David Garlan, Len Bass, Judith Stafford, Robert Nord, James
  Ivers, and Reed Little.
\newblock {\em Documenting software architectures: views and beyond}.
\newblock Pearson Education, 2002.

\bibitem{Hole2016Anti}
Kjell~Jørgen Hole.
\newblock Anti-fragile ict systems.
\newblock 2016.

\bibitem{nguyen2018multi}
Chi~Mai Nguyen, Roberto Sebastiani, Paolo Giorgini, and John Mylopoulos.
\newblock Multi-objective reasoning with constrained goal models.
\newblock {\em Requirements Engineering}, 23(2):189--225, 2018.

\bibitem{darimont1997grail}
Robert Darimont, Emmanuelle Delor, Philippe Massonet, and Axel van Lamsweerde.
\newblock Grail/kaos: an environment for goal-driven requirements engineering.
\newblock In {\em Proceedings of the (19th) international conference on
  software engineering}, pages 612--613. IEEE, 1997.

\bibitem{lamsweerde2003kaos}
A~Lamsweerde.
\newblock Kaos tutorial.
\newblock {\em Cediti, September}, 5, 2003.

\bibitem{garriga2017towards}
Martin Garriga.
\newblock Towards a taxonomy of microservices architectures.
\newblock In {\em International Conference on Software Engineering and Formal
  Methods}, pages 203--218. Springer, 2017.

\bibitem{nygard2018release}
Michael~T Nygard.
\newblock {\em Release it!: design and deploy production-ready software}.
\newblock Pragmatic Bookshelf, 2018.

\bibitem{Aderaldo2017Benchmark}
Carlos~M. Aderaldo, Claus Pahl, and Pooyan Jamshidi.
\newblock Benchmark requirements for microservices architecture research.
\newblock In {\em International Workshop on Establishing the Community-wide
  Infrastructure for Architecture-based Software Engineering}, 2017.

\bibitem{beyer2016site}
Betsy Beyer, Chris Jones, Jennifer Petoff, and Niall~Richard Murphy.
\newblock {\em Site Reliability Engineering: How Google Runs Production
  Systems}.
\newblock " O'Reilly Media, Inc.", 2016.

\end{thebibliography}

\end{document}